%% file: dofxfull.tex
\documentclass[onecolumn, conference]{IEEEtran}

\usepackage{amssymb}
\usepackage{amsmath}
\usepackage{epsfig}
\usepackage{cite}
\usepackage{verbatim}
\usepackage{enumerate}
\usepackage{subfigure}
\usepackage{multicol}
\usepackage{picinpar}
\usepackage{pstricks}
\usepackage{pst-node}

\newcommand{\define}{\stackrel{\triangle}{=}}

\newtheorem{theorem}{\bf Theorem}

\newtheorem{corollary}{\bf Corollary}
\newtheorem{lemma}{\bf Lemma}

\begin{document}
\setcounter{page}{1}
\title{{Degrees of Freedom Region for the \\MIMO $X$ Channel}}
\author{\authorblockN{Syed A. Jafar}
\authorblockA{Electrical Engineering and Computer Science\\
University of California Irvine, \\
Irvine, California, 92697, USA\\
Email: syed@uci.edu\\ \vspace{-1cm}}
\and
\authorblockN{Shlomo Shamai (Shitz)}
\authorblockA{Department of Electrical Engineering \\
Technion-Israel Institute of Technology\\
Technion City, Haifa 32000, Israel\\
Email: sshlomo@ee.technion.ac.il\\ \vspace{-1cm}
}}

\maketitle
\thispagestyle{empty}
\begin{abstract} We provide achievability as well as converse results for the degrees of freedom region of a MIMO $X$ channel, i.e.,  a system with two transmitters, two receivers, each equipped with multiple antennas, where independent messages need to be conveyed over fixed channels from each transmitter to each receiver. The inner and outerbounds on the degrees of freedom region are tight whenever integer degrees of freedom are optimal for each message. With $M=1$ antennas at each node, we find that the total (sum rate) degrees of freedom are bounded above and below as $1 \leq\eta_X^\star \leq \frac{4}{3}$.  If $M>1$ and channel matrices are non-degenerate then the precise degrees of freedom $\eta_X^\star = \frac{4}{3}M$. Thus, the MIMO $X$ channel has non-integer degrees of freedom when $M$ is not a multiple of $3$. Simple zero forcing without dirty paper encoding or successive decoding, suffices to achieve the  $\frac{4}{3}M$ degrees of freedom. The key idea for the achievability of the degrees of freedom is \emph{interference alignment} - i.e., signal spaces are aligned at receivers where they constitute interference while they are separable at receivers where they are desired.   With equal number of antennas at all nodes, we explore the increase in degrees of freedom when some of the messages are made available to a transmitter or receiver in the manner of cognitive radio. With a cognitive transmitter, i.e. with one message shared between transmitters on the MIMO $X$ channel we show that the number of degrees of freedom $\eta = \frac{3}{2}M$ (for $M>1$). The same degrees of freedom are obtained on the MIMO $X$ channel with a cognitive receiver as well, i.e. when one message is made available to its non-intended receiver. In contrast to the $X$ channel result, we show that for the MIMO \emph{interference} channel, the degrees of freedom are not increased even if both the transmitter and the receiver of one user know the other user's message. However, the interference channel can achieve the full $2M$ degrees of freedom if \emph{each} user has either a cognitive transmitter or a cognitive receiver. Lastly, if the channels vary with time/frequency then the $X$ channel with single antennas $(M=1)$ at all nodes has exactly $4/3$ degrees of freedom with no shared messages and exactly $3/2$ degrees of freedom with a cognitive transmitter or a cognitive receiver.
\end{abstract}

\newpage


\section{Introduction}
There is recent interest in the degrees of freedom\footnote{In this paper we use the terms ``multiplexing gain'' and ``degrees of freedom'' interchangeably.} for {distributed} multiple input multiple output (MIMO) communication systems. The distributed MIMO perspective is relevant not only for wireless networks where the nodes are equipped with multiple antennas but also for networks of single antenna nodes which may achieve MIMO behavior through message sharing and collective relaying by clusters of neighboring nodes \cite{Boelcskei_Nabar_Oyman_Paulraj, Morgenshtern_Boelcskei_Nabar, Nabar_Boelcskei_Morgenshtern,Gupta_Kumar_achievable, Borade_Zheng_Gallager, Muller, Kramer_Gastpar_Gupta}. While time, frequency and space all offer degrees of freedom \cite{Poon_Tse, Hanlen_Abhayapala}, spatial dimensions are especially interesting for how they may be accessed with distributed processing. A number of possibilities arise in a wireless network with distributed nodes and with multiple (possibly varying across users) antennas at each transmitter and receiver. One can create non-interfering channels through spatial zero forcing \cite{Spencer_Swindlehurst_Haardt}, i.e. beamforming in the null space of interference signals. Successive decoding and dirty paper coding \cite{Costa} are powerful techniques that can also eliminate interference. The number of interference free dimensions that can be created depends on how the signal vectors may be aligned relative to each other. While the signal space may have potentially as many spatial dimensions as the total number of transmit and receive antennas across all the nodes in the network, optimal signal alignment is a challenging task because the access to these dimensions is restricted by the distributed nature of the network. Some of these restrictions may be circumvented by cooperation among nodes through the sharing and collective relaying of messages. Message sharing, beamforming, zero forcing, successive decoding and dirty paper coding techniques may be combined in many different ways across users, data streams and antennas to establish innerbounds on the degrees of freedom. To determine the \emph{maximum} degrees of freedom one also needs a converse, or an upperbound on the multiplexing gain that is not limited to specific schemes. In this work we provide achievability as well as converse arguments for the degrees of freedom region of a MIMO $X$ channel, i.e.,  a system with two transmitters, two receivers, each equipped with multiple antennas, where independent messages need to be conveyed from each transmitter to each receiver. We also consider the benefits of transmitter side cooperation in the form of shared messages that are available to both transmitters.

Previous work by several researchers \cite{Jafar_dof_int, MadsenIT, Nosratinia-Madsen,Weingarten_Shamai_Kramer} has determined the degrees of freedom for various multiuser MIMO systems. The single user point to point MIMO channel with $M_1$ transmit and $N_1$ receive antennas is known to have $\min(M_1,N_1)$ degrees of freedom \cite{Foschini_Gans,Telatar}. For the two user MIMO multiple access channel (MAC) with $N_1$ receive antennas and $M_1,M_2$ transmit antennas at the two transmitters, the maximum multiplexing gain is $\max(M_1+M_2,N_1)$ \cite{Tse_Viswanath_Zheng}. Thus, the multiplexing gain is the same as the point to point MIMO channel with full cooperation among all transmit antennas. The two user broadcast channel (BC) with $M_1$ transmit antennas and $N_1, N_2$ receive antennas has a maximum multiplexing gain of $\max(M_1,N_1+N_2)$ which is also the same as the point to point MIMO channel obtained with full cooperation between the two receivers \cite{Yu_Cioffi,Viswanath_Tse_BC,Vishwanath_Jindal_Goldsmith}. The multiplexing gain for two user MIMO interference channels is found in \cite{Jafar_dof_int}. It is shown that for a $(M_1,N_1, M_2,N_2)$  MIMO interference channel (i.e. a MIMO interference channel with $M_1, M_2$ antennas at the two transmitters and $N_1,N_2$  antennas at their respective receivers), the maximum multiplexing gain is equal to $\min\left(\scriptstyle{M_1+M_2, N_1+N_2, \max(M_1,N_2),\max(M_2,N_1)}\right)$. \cite{Borade_Zheng_Gallager} considers the degrees of freedom for a multilayer (multiple orthogonal hops) distributed relay network where the source and destination nodes are equipped with $n$ antennas each and there are $n$ single antenna relay nodes at each layer. With only one layer of relay nodes (2 hops) the BC from source to relay nodes and the MAC from the relays to the destination node are concatenated so that $n$ degrees of freedom are achieved inspite of the distributed processing at the intermediate relay nodes. The case of 2-layers (three hops) with, say $n=2$ relay nodes at each hop, is especially interesting, as the intermediate hop takes place over an interference channel with single antenna nodes.  While the two user interference channel with single antenna nodes has only one degree of freedom by itself, it is able to deliver $2$ degrees of freedom when used as an intermediate stage between a $2$ antenna source and a $2$ antenna destination \cite{Borade_Zheng_Gallager}. The key is an amplify and forward scheme where the relay nodes, instead of trying to decode the messages, simply scale and forward their received signals. \cite{Boelcskei_Nabar_Oyman_Paulraj, Morgenshtern_Boelcskei_Nabar, Nabar_Boelcskei_Morgenshtern} consider end to end channel orthogonalization with distributed sources, relays and destination nodes and determine the capacity scaling behavior with the number of relay nodes. It is shown that distributed orthogonalization can be obtained even with synchronization errors if a minimum amount of coherence at the relays can be sustained. Degrees of freedom for linear interference networks with local side-information are explored in \cite{Lapidoth_Shamai_Wigger_IN} and cognitive message sharing is found to improve the degrees of freedom for certain structured channel matrices.

The MIMO MAC and BC channels show that there is no loss in degrees of freedom even if antennas are distributed among users at one end (either transmitters or receivers) making joint signal processing infeasible, as long as joint signal processing is possible at the other end of the communication link. The multiple hop example of \cite{Borade_Zheng_Gallager}, described above, shows that there is no loss of degrees of freedom even with distributed antennas at both ends of a communication hop (an interference channel) as long as the distributed antenna stages are only intermediate hops and joint processing can take place at the source and destination terminals that are equipped with multiple antennas. However the MIMO interference channel (IC) shows that if antennas are distributed at both ends then the degrees of freedom can be severely limited. For example, consider a MIMO MAC or BC where the total number of transmit antennas is $n$ and the total number of receive antennas is also $n$. Regardless of how the transmit or receive antennas are distributed among two users, both the multiple access channel and the broadcast channel are capable of achieving the maximum multiplexing gain of $n$. However, consider the $(1, n-1, n-1, 1)$ interference channel which also has a total of $n$ transmit antennas and $n$ receive antennas, but the maximum multiplexing gain of this interference channel is only $1$. Thus, distributed processing at both ends severely limits the degrees of freedom. 

Researchers have also explored if some of the loss in degrees of freedom from distributed processing can be recovered by allowing noisy communication links between distributed transmitters or distributed receivers so that they can cooperate and share information. These investigations have primarily focused on single antenna nodes. The two user interference channel with single antennas at all nodes is considered by Host-Madsen \cite{MadsenIT}. It is shown that the maximum multiplexing gain is only equal to one even if cooperation between the two transmitters or the two receivers is allowed via a noisy communication link.  Host-Madsen and Nosratinia \cite{Nosratinia-Madsen} show that even if noisy communication links are introduced between the two transmitters as well as between the two receivers the highest multiplexing gain achievable is equal to one. 

Another form of cooperation between transmitters is to allow message sharing, i.e. one transmitter's message is made available non-causally to the other transmitter. This channel is called the ``cognitive radio'' channel in \cite{Devroye_Mitran_Tarokh, Devroye_Mitran_Tarokh_Mag} and its capacity region was determined under the assumption of weak interference in \cite{Wu_Vishwanath_Arapostathis,Jovicic_Viswanath}. With single antennas at all nodes, it was shown recently in \cite{Devroye_Sharif} that even this form of unidirectional (from one transmitter to another) noiseless cooperation does not produce any gain in the degrees of freedom. These results are somewhat surprising as it can be shown that with ideal cooperation between transmitters (broadcast channel) or with ideal cooperation between receivers (multiple access channel) the maximum multiplexing gain is equal to 2.

It is interesting to note that for all the cases discussed above, spatial zero forcing suffices to achieve all the available degrees of freedom. It is shown in \cite{Jafar_dof_int} that all the degrees of freedom on the MIMO interference channel, the MIMO broadcast channel as well as the MIMO multiple access channel can be achieved purely by spatial zero forcing. All these results may be seen as negative results because they suggest that for multiplexing gain in distributed MIMO channels, there is nothing more beyond spatial zero forcing.
\subsection{The MIMO $X$ channel}
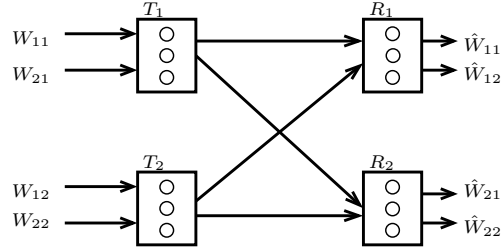
\begin{figure}[h]
\centerline{\input{xchannel.pstex_t}}
\caption{MIMO $X$ Channel}\label{fig:Xchannel}
\end{figure}
The MIMO $X$ channel is shown in Figure \ref{fig:Xchannel} and is described by the input output equations:
\begin{eqnarray*}
{\bf Y}^{[1]}&=&{\bf H}^{[11]}{\bf X}^{[1]}+{\bf H}^{[12]}{\bf X}^{[2]}+{\bf N}^{[1]}\\
{\bf Y}^{[2]}&=&{\bf H}^{[21]}{\bf X}^{[1]}+{\bf H}^{[22]}{\bf X}^{[2]}+{\bf N}^{[2]}
\end{eqnarray*}
where ${\bf Y}^{[1]}$ is the $N_1\times 1$ output vector at receiver 1, ${\bf Y}^{[2]}$ is the $N_2\times 1$ output vector at receiver 2, ${\bf N}^{[1]}$ is the $N_1\times 1$ additive white Gaussian noise (AWGN) vector at receiver 1, ${\bf N}^{[2]}$ is the $N_2\times 1$ AWGN vector at receiver $2$, ${\bf X}^{[1]}$ is the $M_1\times 1$ input vector at transmitter $1$, ${\bf X}^{[2]}$ is the $M_2\times 1$ input vector at transmitter 2, ${\bf H}^{[11]}$ is the $N_1\times M_1$ channel matrix between transmitter 1 and receiver 1, ${\bf H}^{[22]}$ is the $N_2\times M_2$ channel matrix between transmitter 2 and receiver 2, ${\bf H}^{[12]}$ is the $N_1\times M_2$ channel matrix between transmitter 2 and receiver 1, and  ${\bf H}^{[21]}$ is the $N_2\times M_1$ channel matrix between transmitter 1 and receiver 2.  As shown in Figure \ref{fig:Xchannel} there are four independent messages in the MIMO $X$ channel: $W_{11}, W_{12}, W_{21}, W_{22}$ where $W_{ij}$ represents a message from transmitter $j$ to receiver $i$. 

We assume the channel matrices are generated from a continuous probability distribution so that, almost surely, any matrix composed of channel coefficients will have rank equal to the minimum of the number of its rows and columns. Perfect knowledge of all channel coefficients is available to  all transmitters and receivers. With the exception of Section \ref{section:normalized}, we assume throughout that the values of the channel coefficients  are fixed throughout the duration of communication. The implications of time/frequency selective fading are briefly discussed in Section \ref{section:normalized}.

The power at each transmitter is assumed to be equal to $\rho$. We indicate the size of the message set by $|W_{ij}(\rho)|$. For codewords spanning $n$ channel uses, the rates $R_{ij}(\rho)=\frac{\log|W_{ij}(\rho)|}{n}$ are achievable if the probability of error for \emph{all} messages can be simultaneously made arbitrarily small by choosing an appropriately large $n$. For rate functions $R_{ij}(\rho)$ we define the degrees of freedom
\begin{equation}
d_{ij}=\lim_{\rho\rightarrow\infty}\frac{R_{ij}(\rho)}{\log(\rho)}.
\end{equation}

We define the \emph{degrees of freedom region} for the MIMO $X$ channel as:
\begin{eqnarray*}
\mathcal{D}^X\define &&\left\{(d_{11},d_{12},d_{21},d_{22}): d_{ij}=\lim_{\rho\rightarrow\infty}\frac{R_{ij}(\rho)}{\log(\rho)},~~  \mbox{Prob}(\hat W_{ij}\neq W_{ij}(\rho)) \rightarrow 0~~\mbox{as}~~ n\rightarrow\infty~~\forall i,j\in\{1,2\} \right\}
\end{eqnarray*}
and the total degrees of freedom $\eta_X^\star$
\begin{eqnarray*}
\eta_X^\star&\define & \max_{\mathcal{D}^X}(d_{11}+d_{12}+d_{21}+d_{22})
\end{eqnarray*}


The MIMO $X$ channel is especially interesting because it is generalizes the interference channel to allow an independent message from each transmitter to each receiver.  An interesting coding scheme is recently proposed by Maddah-Ali, Motahari and Khandani in \cite{MMK} for the two user MIMO $X$ channel with three antennas at all nodes. Just as the MIMO $X$ channel combines elements of the MIMO broadcast channel, the MIMO multiple access channel and the MIMO interference channel into one channel model, the MMK scheme naturally combines dirty paper coding, successive decoding and zero forcing elements into an elegant coding scheme tailored for the MIMO $X$ channel. The results of \cite{Jafar_dof_int} establish that with $3$ antennas at all nodes, the maximum multiplexing gain for each of the MIMO IC, MAC and BC channels contained within the $X$ channel is $3$. However, for the MIMO $X$ channel with $3$ antennas at all nodes, the MMK scheme is able to achieve $4$ degrees of freedom. The MMK scheme also extends easily to achieve $4M$ degrees of freedom on the MIMO $X$ channel with $3M$ antennas at each node. Thus, the results of \cite{MMK} show that the degrees of freedom on the MIMO $X$ channel strictly surpass what is achievable on the interference, multiple access and broadcast components individually. 

Several interesting questions arise for the MIMO $X$ channel. First, we need an outerbound to determine what is the maximum multiplexing gain for the MIMO $X$ channel, and in particular, if the MMK scheme is optimal. Second, we note that neither dirty paper coding nor successive decoding have ever been found to be necessary to achieve the full degrees of freedom on any multiuser MIMO channel with perfect channel knowledge. Zero forcing suffices to achieve all degrees of freedom on the MIMO MAC, BC, and interference channels. So the natural question is whether zero forcing also suffices to achieve all the degrees of freedom for the MIMO $X$ channel. Third, if the factor of $4/3M$ suggested by the results of \cite{MMK} is found to be optimal, it would lead to noninteger values for the degrees of freedom when $M$ is not an integer multiple of $3$. This is of fundamental interest because there are no known results for the optimality of non-integer degrees of freedom for any non-degenerate wireless network with perfect channel knowledge\footnote{Degrees of freedom with channel uncertainty have been explored in \cite{Lapidoth,Lapidoth_Shamai_Wigger_BC, Lapidoth_Shamai_collapse, Jindal_BCFB, Sharif_Hassibi}.}. Finally, while the interference channel does not seem to benefit from cooperation through noisy channels between transmitters and receivers, it is not known if shared messages (in the manner of cognitive radio \cite{Devroye_Mitran_Tarokh}) can improve the degrees of freedom on the MIMO $X$ and interference channels. These are the open questions that we answer in this work. 

\subsection{Overview of Results}
We provide achievability and converse results for the degrees of freedom region for all messages on the MIMO $X$ channel. The inner and outerbounds are characterized by the same set of linear inequalities on the degrees of freedom for the four messages, with the difference that the outerbound allows all real values while the innerbound is restricted to the convex hull of integer values for the degrees of freedom. We also explicitly solve these linear inequalities to characterize the maximum degrees of freedom $\eta_X^\star$ for the sum rate of the MIMO $X$ channel. We show that at least three fourths of the maximum multiplexing gain  of the MIMO $X$ channel can be achieved by at least one of the MAC, BC and IC components. For equal number of antennas at all nodes $M>1$ we show that the MIMO $X$ channel has precisely $4/3M$ degrees of freedom. Thus we establish that the MIMO $X$ channel has noninteger degrees of freedom when $M>1$ and $M$ is not a multiple of $3$. For the $X$ channel with a single antenna at each node, $1\leq\eta_X^\star\leq\frac{4}{3}$. 

\begin{figure}[h]
\centerline{\input{xalign.pstex_t}}
\caption{Interference Alignment on the MIMO $X$ Channel}\label{fig:xalign}
\end{figure}
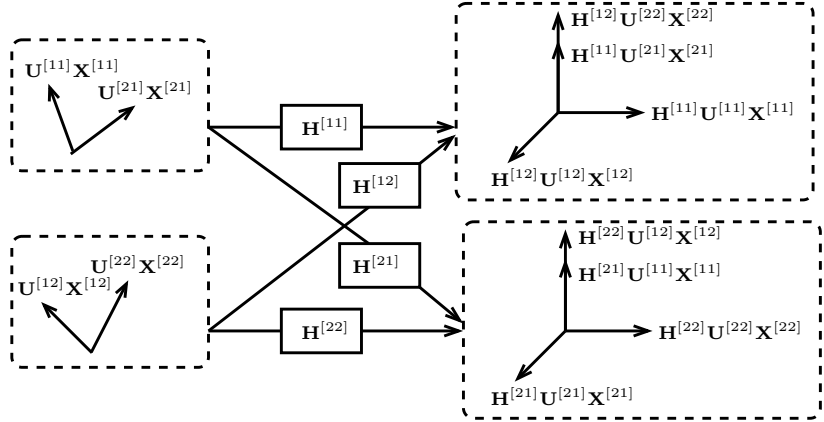

Several interesting observations can be made regarding the schemes used in this work to establish the achievable degrees of freedom for the MIMO $X$ channel. First, these schemes do not require dirty paper coding or successive decoding. Instead, as with the MIMO MAC, BC and interference channels, the optimal achievability schemes are based on simple zero forcing. The distinguishing feature of the MIMO $X$ channel is the concept of interference alignment illustrated in Figure \ref{fig:xalign}.

As shown in the figure, transmitter $1$ transmits independent codewords ${\bf X}^{[11]},{\bf X}^{[21]}$ for messages $W_{11}, W_{21}$ along beamforming directions ${\bf U}^{[11]}, {\bf U}^{[21]}$ while transmitter $2$ sends independent codewords  ${\bf X}^{[12]},{\bf X}^{[22]}$ for messages $W_{12}, W_{22}$ along beamforming directions ${\bf U}^{[12]}, {\bf U}^{[22]}$, respectively. The transmit vectors undergo the linear transformations represented by the channel matrices ${\bf H}^{[ij]}$. Interference alignment refers to the careful choice of beamforming directions in such a manner that the desired signals are separable at their respective receivers while the interference signals are aligned, i.e., the interference vectors cast \emph{overlapping shadows}. In Figure \ref{fig:xalign}, the desired signal vectors ${\bf H}^{[11]}{\bf U}^{[11]}{\bf X}^{[11]}$ and  ${\bf H}^{[12]}{\bf U}^{[12]}{\bf X}^{[12]}$ are linearly independent while the interference vectors ${\bf H}^{[12]}{\bf U}^{[22]}{\bf X}^{[22]}$ and ${\bf H}^{[11]}{\bf U}^{[21]}{\bf X}^{[21]}$ are linearly dependent so that they occupy the same spatial dimensions as seen by receiver 1. A similar alignment occurs at receiver $2$ as well. The advantage of interference alignment is that zero forcing one interference signal automatically zero forces \emph{both} interference signals. In other words, discarding the dimensions spanned by one interference signal also eliminates the other interference signal, so that the interference free dimensions available for desired signals are maximized. Interference alignment is pointed out as a useful idea for the MIMO $X$ channel by Maddah-Ali, Motahari and Khandani in \cite{MMK}. Interference alignment is found to be particularly useful for the compound broadcast channel in \cite{Weingarten_Shamai_Kramer}.

Another distinguishing feature of the MIMO $X$ channel is that it can have noninteger degrees of freedom. To the best of our knowledge the MIMO $X$ channel is the first example of a multiuser communication scenario with non-degenerate channels and full channel knowledge where noninteger degrees of freedom are optimal. For example, the point to point MIMO channel, and the MIMO MAC, BC and interference channels all have integer degrees of freedom. With equal number ($M>1$) of antennas  at all nodes, achievability of noninteger degrees of freedom is established by interference alignment and zero forcing over the 3-symbol extension of the channel. While the extended channel idea does not help with the $M=1$ case, for $M>1$ and non-degenerate channel matrices it allows us enough dimensions to construct and align the signal vectors as shown in Figure \ref{fig:xalign}. For $M=1$ we are also able to achieve the full $4/3$ degrees of freedom if the channel coefficients are time/frequency selective.

\begin{figure}[h]
\centerline{\input{intcogtxrx1.pstex_t}}
\caption{Cognitive MIMO Interference Channels with $\eta^\star=M$.}\label{fig:cogint}
\end{figure}
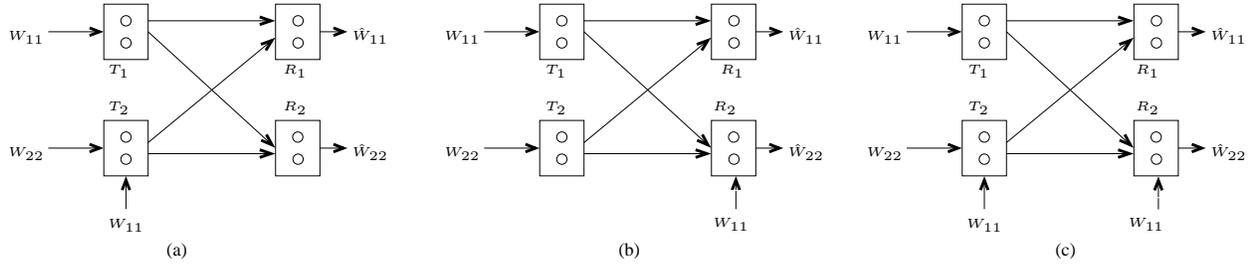

Next we explore the impact of shared messages on the degrees of freedom for the MIMO $X$ channel and its special case, the MIMO interference channel. For simplicity we consider the case where all nodes have equal number of antennas $M$. First, consider the MIMO interference channel with $M$ antennas at all nodes. We show that the total number of degrees of freedom $\eta^\star$ for the MIMO interference channel is not increased by sharing one user's message with another user's transmitter, receiver or both (as shown in Fig. \ref{fig:cogint}(a), Fig. \ref{fig:cogint}(b), Fig. \ref{fig:cogint}(c), respectively). In all these cases the degrees of freedom are the same as without any cognitive transmitters or receivers, $\eta^\star=M$. 

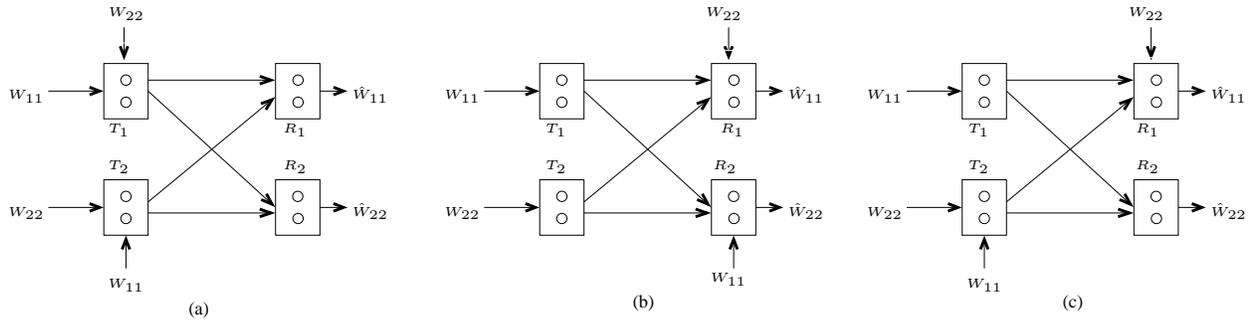
\begin{figure}[h]
\centerline{\input{intcogfull.pstex_t}}
\caption{Cognitive MIMO Interference Channels with $\eta^\star=2M$.}\label{fig:cogintfull}
\end{figure}

However, the interference channel can achieve the full $2M$ degrees of freedomas if both users have cognitive transmitters, or they both have cognitive receivers, or one user has a cognitive transmitter while the other user has a cognitive receiver (as shown in Fig. \ref{fig:cogintfull}(a), Fig. \ref{fig:cogintfull}(b), Fig. \ref{fig:cogintfull}(c), respectively).

In contrast to the MIMO interference channel, the MIMO $X$ channel does benefit from cognitive sharing of even a single message. For $M>1$, with any one message (e.g. $W_{11}$) made available to the other transmitter (transmitter $2$) or its unintended receiver (receiver $2$) the number of degrees of freedom on the MIMO $X$ channel is $\frac{3}{2}M$. 
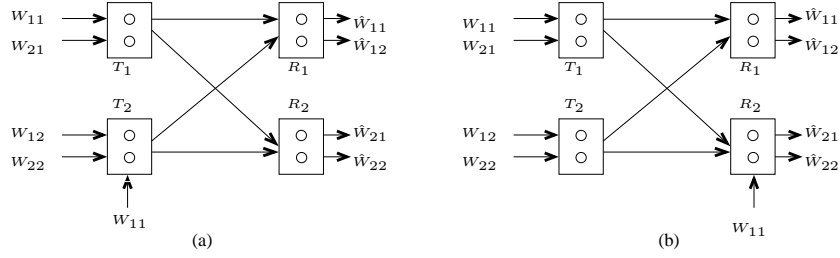
\begin{figure}[h]
\centerline{\input{cogxfull.pstex_t}}
\caption{MIMO $X$ Channels with (a) Cognitive Transmitter and (b) Cognitive Receiver. In both cases, $\eta^\star=\frac{3}{2}M$ for $M>1$.}\label{fig:cogxfull}
\end{figure}

It is interesting to note that the degrees of freedom for the MIMO $X$ channel increase according to $\frac{4}{3}M$ for no shared messages$\rightarrow \frac{3}{2}M$ for one shared message$\rightarrow \frac{2}{1}M$ for two shared messages (provided the two shared messages are not intended for the same receiver). The symmetry of the results for degrees of freedom with cognitive transmitters and cognitive receivers is also interesting as it points to a reciprocity relationship between the transmitter and receiver side cognitive cooperation.

{\it Notation:} co(A) is the convex hull of the set A. $\sigma_{\max({\bf H})}$ is the principal singular value of the matrix ${\bf H}$. $(x)^+$ represents the function $\max(x,0)$. $\mathbb{R}_+^n$ and $\mathbb{Z}^n_+$ represent the set of n-tuples of non-negative real numbers and integers respectively.

\section{The MIMO $Z$ and Interference Channels}
\begin{figure}[h]
\centerline{\input{zint.pstex_t}}
\caption{MIMO $Z$ Channel and MIMO Interference Channel}\label{fig:Zint}
\end{figure}
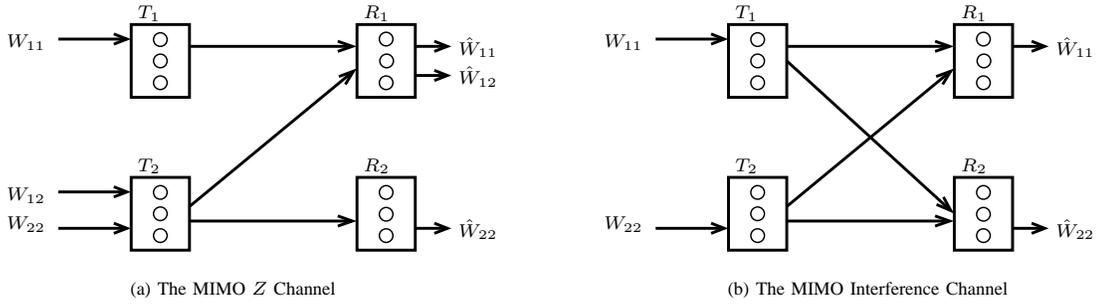

The MIMO interference channel and the MIMO $Z$ channel are depicted in Figure \ref{fig:Zint}. The interference channel and the $Z$ channel are characterized by the same input output equations as the $X$ channel. The distinction between the $X$ and interference channels is made purely based on the constraints on the messages. The $X$ channel is the most general case where each transmitter has an independent message for each receiver, for a total of 4 independent messages. The interference channel $I(11,22)$ is a special case of the $X$ channel with the constraint $W_{12}=W_{21}=\phi$, i.e. there is no message to be communicated from transmitter $1$ to receiver $2$ or from transmitter $2$ to receiver $1$. The $X$ channel contains two interference channels: $I(11,22)$ and $I(12,21)$. 

The $Z$ channel as depicted in Figure \ref{fig:Zint} also corresponds to the $X$ channel with the added constraint that $W_{21}=\phi$ and ${\bf H}_{21}={\bf 0}$. Thus, there is no message or channel from transmitter $1$ to receiver $2$. The $X$ channel is associated with $4$ different $Z$ channels, depending on which message and its corresponding channel are eliminated. We denote these $Z$ channels as $Z(11), Z(12), Z(21), Z(22)$, so that $Z(ij)$ corresponds to the $Z$ channel obtained from the $X$ channel by setting $W_{ij}=\phi$ and ${\bf H}_{ij}={\bf 0}, \forall i,j\in\{1,2\}.$

Similar to the $X$ channel, the achievable rates and the degrees of freedom can be defined for the messages in the $Z$ channel and the interference channel so that:
\begin{eqnarray}
\eta_{Z(21)}^\star&\triangleq &\max_{\mathcal{D}^{Z(21)}}(d_{11}+d_{12}+d_{22}),\\
\eta_{I(11,22)}^\star&\triangleq & \max_{\mathcal{D}^{I(11,22)}}(d_{11}+d_{22}).
\end{eqnarray}
In this work, our interest in the interference and $Z$ channels is limited to how they can be used to derive outerbounds for the degrees of freedom on the MIMO $X$ channel. The following lemma states the relationship between the degrees of freedom on these channels.
\begin{lemma}\label{lemma:xzint}
\begin{eqnarray}
\max_{\mathcal{D}^X}(d_{11}+d_{22})&\leq &\max{\mathcal{D}^{I(11,22)}} (d_{11}+d_{22})=\eta_{I(11,22)}^\star\\
\max_{\mathcal{D}^X}(d_{11}+d_{12}+d_{22})&\leq &\max{\mathcal{D}^{Z(21)}} (d_{11}+d_{12}+d_{22})=\eta_{Z(21)}^\star
\end{eqnarray}
\end{lemma}
\begin{proof}
The first bound is straightforward because the interference channel is obtained by eliminating messages $W_{12}$ and $W_{21}$ from the $X$ channel. Any coding scheme for the $X$ channel can be used on the interference channel by picking $W_{12}$ and $W_{21}$ as known sequences shared beforehand between all transmitters and receivers as a part of the codebook, rather than messages that are unknown apriori. Therefore if $d_{11}$ and $d_{22}$ are achievable on the $X$ channel, then they are also achievable on the interference channel.

For the second bound, suppose we have a coding scheme that is able to achieve $d_{11},d_{12},d_{22}$ on the $X$ channel. Now suppose, in place of message $W_{21}$ we use a known sequence that is available to all transmitters and receivers apriori. Also, a genie provides $W_{11}$ to receiver $2$. Thus, receiver $2$ knows all the information available to transmitter 1 and can subtract transmitter 1's signal from its received signal. This is equivalent to ${\bf H}^{[21]}={\bf 0}$, so the resulting $X$ channel becomes identical to the $Z$ channel of Figure \ref{fig:Zint}. However, neither setting $W_{21}$ to a known sequence, nor the genie information to receiver $2$ can deteriorate the performance of the coding scheme. Therefore the same degrees of freedom $d_{11}^Z=d_{11}, d_{12}^Z=d_{12}, d_{22}^Z=d_{22}$ are achievable on the $Z$ channel as well.
\end{proof}

Lemma \ref{lemma:xzint} is useful because the degrees of freedom for the MIMO interference channel are already known and outerbounds on the degrees of freedom for the MIMO $Z$ channel can be obtained as we show in the next section.

\subsection{Outerbounds on the degrees of freedom for the MIMO $Z$ channel}
In order to obtain outerbounds for the $X$ channel we start with the $Z$ channel and derive an upperbound  on its sum rate in terms of the sum rate of a corresponding multiple access channel (MAC).
\begin{theorem}\label{theorem:outer}
If $N_1\geq M_2$, then for the $Z(21)$ channel described above, the sum capacity is bounded above by that of the corresponding MAC channel from transmitters $1$ and $2$ to receiver $1$ and with additive noise ${\bf N^{[1]}}\sim\mathcal{N}({\bf 0, I_{N_1}})$ modified to ${\bf N^{(1)'}}\sim\mathcal{N}({\bf 0, K^{'}})$ where
\begin{eqnarray*}
{\bf K^{'}}&=& \scriptstyle{{\bf I_{N_1}}- {\bf H}^{[12]}\left({\bf 
H}^{[12]\dagger}{\bf H}^{[12]}\right)^{-1}{\bf H}^{[12]\dagger}+\alpha{\bf 
H}^{[12]}{\bf H}^{[12]\dagger}},\\
\alpha & = & \min\left(\frac{1}{\sigma^2_{\max}({\bf H}^{[12]})},\frac{1}{\sigma^2_{\max}({\bf H}^{[22]})}\right).
\end{eqnarray*}
\end{theorem}
\begin{proof}
The proof is similar to the proof for Theorem $1$ in \cite{Jafar_dof_int}. Instead of repeating the details we provide a sketch of the proof. In the original $Z(21)$ channel, receiver 1 must decode message $W_{11}$ and $W_{12}$. Since $W_{11}$ is the only message sent from transmitter $1$, decoding $W_{11}$ allows receiver $1$ to eliminate transmitter $1$'s contribution to the received signal. By reducing the noise at receiver $1$ we make receiver $1$ less noisy than receiver $2$. This can be done only if $M_2\leq N_1$ because otherwise, no matter how small the noise, it is possible for transmitter $2$ to transmit to receiver $2$ along the null space of its channel to receiver $1$. Since receiver $2$ is able to decode $W_{22}$ and receiver $1$ has a less noisy version of receiver $2$'s output, it can also decode $W_{22}$. Thus, receiver $1$ in the  resulting multiple access channel is able to decode all three messages $W_{11}, W_{12}, W_{22}$ and its sum-rate cannot be smaller than the original $Z(21)$ channel. 
\end{proof}

The following corollary is a direct consequence of Theorem $1$.
\begin{corollary} \label{cor:z}
$\eta_{Z(21)}^\star = \max_{\mathcal{D}^{Z(21)}}(d_{11}+d_{12}+d_{22})\leq \max(N_1,M_2)$.
\end{corollary}
\begin{proof}
If $N_1\geq M_2$ then from Theorem \ref{theorem:outer} the sum capacity is bounded by the MAC with $N_1$ receive antennas. If $N_1<M_2$ let us add more antennas to receiver $1$ so that the total number of antennas at receiver 1 is equal to $M_2$.  Additional receive antennas cannot hurt so the converse argument is not violated. The sum capacity of the resulting $Z(21)$ channel is bounded above by the MAC with $M_2$ receive antennas. The  multiplexing gain on a MAC  cannot be more than the total number of receive antennas. Therefore, in all cases $\eta_{Z(21)}^\star\leq \max(N_1,M_2)$. 
\end{proof}
\section{Degrees of freedom Region for the MIMO $X$ channel}
\subsection{Outerbound}
\begin{theorem}$\mathcal{D}^X\subset\mathcal{D}^X_{out}$ where the outerbound on the degrees of freedom region is defined as follows.
\begin{eqnarray*}
\mathcal{D}^X_{out}&\define&\left\{(d_{11},d_{12},d_{21},d_{22})\in \mathbb{R}^4_+:\right.\\
&& d_{11}+d_{12}+d_{21}\leq\max(N_1,M_1)\\
&& d_{11}+d_{12}+d_{22}\leq\max(N_1,M_2)\\
&& d_{11}+d_{21}+d_{22}\leq\max(N_2,M_1)\\
&& d_{12}+d_{21}+d_{22}\leq\max(N_2,M_2)\\
&& d_{11}+d_{12} \leq N_1\\
&& d_{21}+d_{22} \leq N_2\\
&& d_{11}+d_{21} \leq M_1\\
&&\left. d_{12}+d_{22} \leq M_2\right\}
\end{eqnarray*}
\end{theorem}
\begin{proof}
From the $X$ channel we can form $4$ different $Z$ channels by eliminating one of the $4$ messages and setting the corresponding channel to zero. Combining results of Lemma \ref{lemma:xzint} and Corollary \ref{cor:z} we have
\begin{eqnarray*}
\max_{\mathcal{D}^X}(d_{11}+d_{12}+d_{21})\leq\eta^\star_{Z(22)} \leq \max(N_1, M_1)\\
\max_{\mathcal{D}^X}(d_{11}+d_{12}+d_{22})\leq\eta^\star_{Z(21)}\leq \max(N_1, M_2)\\
\max_{\mathcal{D}^X}(d_{11}+d_{21}+d_{22})\leq\eta^\star_{Z(12)}\leq \max(N_2, M_1)\\
\max_{\mathcal{D}^X}(d_{12}+d_{21}+d_{22})\leq\eta^\star_{Z(11)} \leq \max(N_2, M_2)
\end{eqnarray*}
The last four conditions represent straightforward outerbounds from the multiple access and broadcast channels contained in the MIMO $X$ channel.
\end{proof}
Note that the outerbound allows all real non-negative values for $d_{ij}$ that satisfy the 8 inequalities. The boundary values of $d_{ij}$, e.g., those that maximize $\eta_X$ may not be integers. This is the main distinction between the outerbound and the innerbound to be presented next. 

\subsection{Innerbound}
\begin{theorem}\label{theorem:innerbound}
$\mathcal{D}^X\supset \mathcal{D}^X_{in}\define \mbox{co}\left(\mathcal{D}^X_{out}\cap  \mathbb{Z}^4_+\right)$.
\end{theorem}
\begin{proof}
We provide a constructive achievability proof for Theorem \ref{theorem:innerbound}. The transmitted signals ${\bf X}^{[1]}$ and ${\bf X}^{[2]}$ are chosen as:
\begin{eqnarray*}
{\bf X}^{[1]}&=&\sum_{i=1}^{d_{11}}{\bf v}^{[11]}_ix^{[11]}_i+\sum_{i=1}^{d_{21}}{\bf v}^{[21]}_ix^{[21]}_i\\
{\bf X}^{[2]}&=&\sum_{i=1}^{d_{12}}{\bf v}^{[12]}_ix^{[12]}_i+\sum_{i=1}^{d_{22}}{\bf v}^{[22]}_ix^{[22]}_i\\
\end{eqnarray*}
where $x^{[jk]}_i$ represents the $i^{th}$ input used to transmit the codeword for message $W_{jk}$. 

The transmit direction vectors ${\bf v}^{[21]}_1, \cdots, {\bf v}^{[21]}_{d_{21}}, {\bf v}^{[22]}_1, \cdots, {\bf v}^{[22]}_{d_{22}}$  are selected from the following formulation of the null space of the concatenated channel matrix ${\bf H}\triangleq[{\bf H}^{[11]}~~{\bf H}^{[12]}]$. 
\begin{eqnarray}
\underbrace{[{\bf H}^{[11]}~~{\bf H}^{[12]}]}_{N_1\times(M_1+M_2)}
\underbrace{
\left[
\begin{array}{ccc|ccc|ccc}
|& \cdots & | & | &\cdots&|&|& \cdots & |\\
{\bf v}^{[21]}_1&\cdots &{\bf v}^{[21]}_{r_1}&{\bf 0}&\cdots&{\bf 0}&{\bf v}^{[21]}_{r_1+1}&\cdots & {\bf v}^{[21]}_{r_1+r}\\
|& \cdots & | & | &\cdots&|&|& \cdots & |\\\hline
|& \cdots & | & | &\cdots&|&|& \cdots & |\\
{\bf 0}&\cdots&{\bf 0}&{\bf v}^{[22]}_1&\cdots &{\bf v}^{[22]}_{r_2}&{\bf v}^{[22]}_{r_2+1}&\cdots & {\bf v}^{[22]}_{r_2+r} \\
|& \cdots & | & | &\cdots&|&|& \cdots & |
\end{array}
\right]
}_{(M_1+M_2)\times(M_1+M_2-N_1)^+~~ \mbox{matrix}~{\bf V}~\mbox{with orthonormal columns}}
=\underbrace{
\left[
\begin{array}{ccc}
| & \cdots & | \\
{\bf 0}&\cdots&{\bf 0} \\
| & \cdots & | 
\end{array}
\right]
}_{N_1\times (r_1+r_2+r)}\label{eq:nullspace}
\end{eqnarray}
where
\begin{eqnarray}
r_1&=&(M_1-N_1)^+\\
r_2&=&(M_2-N_1)^+\\
r &=&(M_1+M_2-N_1)^+-r_1-r_2
\end{eqnarray}
Here, ${\bf v}^{[21]}_1, \cdots, {\bf v}^{[21]}_{r_1}$ are the orthonormal basis vectors for the null space of ${\bf H}^{[11]}$. Similarly, ${\bf v}^{[22]}_1, \cdots, {\bf v}^{[22]}_{r_2}$ are the orthonormal basis vectors for the null space of ${\bf H}^{[12]}$. The remaining $r$ column vectors of {\bf V} are the rest of the null space basis vectors for the concatenated matrix ${\bf H}=[{\bf H}^{[11]}~~{\bf H}^{[12]}]$. The product of {\bf H} with these $r_1+r_2+r$ vectors produces all zeros as indicated by the $N_1\times(r_1+r_2+r)$ submatrix of all zeros on the right hand side (RHS) of equation (\ref{eq:nullspace}).

Note that ${\bf v}^{[21]}_i$ and ${\bf v}^{[22]}_j$ are transmit direction vectors for the messages intended for receiver $2$. The above construction chooses these vectors from the null space of receiver $1$'s channel matrices to allow as much zero forcing of interference as possible. Therefore the choice of the first $r_1$ vectors for ${\bf v}^{[21]}_i$ and the first $r_2$ vectors for ${\bf v}^{[22]}_j$ is straightforward. The choice of the next $r$ transmit vectors is interesting because it aligns the interference spaces of the two messages at receiver $1$. Note that, for $i\in\{1,\cdots,r\}$,
\begin{equation}
{\bf H}^{[11]}{\bf v}^{[21]}_{r_1+i}=-{\bf H}^{[12]}{\bf v}^{[22]}_{r_2+i}.
\end{equation}
Thus at receiver 1 the two interference vectors,
\begin{equation}
{\bf H}^{[11]}{\bf v}^{[21]}_{r_1+i}x^{[21]}_{r_1+i}+{\bf H}^{[12]}{\bf v}^{[22]}_{r_2+i}x^{[22]}_{r_2+i}\label{eq:shadow}
\end{equation}
spans only a one dimensional space, and the interference from both these signals can be discarded with the loss of only one dimension at receiver $1$. However, at the desired receiver (receiver 2), the two signal vectors
\begin{equation}
{\bf H}^{[21]}{\bf v}^{[21]}_{r_1+i}x^{[21]}_{r_1+i}+{\bf H}^{[22]}{\bf v}^{[22]}_{r_2+i}x^{[22]}_{r_2+i}
\end{equation}
almost surely span a two dimensional space.

The above construction only specifies ${\bf v}^{[21]}_1,\cdots,{\bf v}^{[21]}_{r+r_1}$ and ${\bf v}^{[22]}_1,\cdots,{\bf v}^{[22]}_{r+r_2}$. The remaining vectors ${\bf v}^{[21]}_{r+r_1+1},\cdots,{\bf v}^{[21]}_{d_{21}}$ and ${\bf v}^{[22]}_{r+r_2+1},\cdots,{\bf v}^{[22]}_{d_{22}}$ can be picked randomly, e.g. according to an isotropic distribution so that they are linearly independent with probability one. Switching indices $1$ and $2$ a similar construction is then applied to pick transmit directions ${\bf v}^{[12]}_i$ and ${\bf v}^{[11]}_j$ as well.

Following the above construction, the vectors ${\bf v}^{[21]}_1, \cdots, {\bf v}^{[21]}_{d_{11}}$ are linearly independent as long as $d_{21}\leq M_1$. Notice that while vectors ${\bf v}^{[21]}_i$ are derived from channel matrices ${\bf H}^{[11]}, {\bf H}^{[12]}$, the vectors  ${\bf v}^{[11]}_j$ are derived from independent channel matrices ${\bf H}^{[21]}, {\bf H}^{[22]}$. Thus, all the signal vectors generated at transmitter 1 are linearly independent with probability one, if:
\begin{equation}
M_1\geq d_{11}+d_{21}.
\end{equation}
Similarly, all signal vectors generated at transmitter 2, i.e. all ${\bf v}^{[12]}_i, {\bf v}^{[22]}_j$ are linearly independent with probability one, if:
\begin{equation}
M_2\geq d_{12}+d_{22}.
\end{equation}
Both these conditions appear explicitly in the definition of the set $\mathcal{D}^X_{in}$. Therefore, all input signal vectors are linearly independent.

The achievability of $(d_{11}, d_{12}, d_{21}, d_{22})$ is now determined by the receiver's ability to obtain enough interference free dimensions for its desired signals. Consider receiver 1. The desired messages are $W_{11}$ and $W_{12}$. The desired signals are transmitted along $d_{11}$ and $d_{12}$ linearly independent directions by transmitters 1 and 2 respectively. Out of the $N_1$ dimensional signal space observed by receiver 1, suppose the interference signal spans $d_I$ dimensions. Then, $d_{11}$ and $d_{12}$ are achievable provided,
\begin{eqnarray}
N_1\geq d_{11}+d_{12}+d_I.\label{eq:condition}
\end{eqnarray}
If the above relationship holds then receiver 1 can suppress interference by discarding the $d_I$ dimensions that contain interference and the remaining $N_1-d_I$ dimensions are enough to achieve $d_{11}+d_{12}$ degrees of freedom on the desired signals.

The received signal,
\begin{eqnarray*}
{\bf Y}^{[1]}&=&\sum_{i=1}^{d_{11}}{\bf H}^{[11]}{\bf v}^{[11]}_ix^{[11]}_i+\sum_{i=1}^{d_{21}}{\bf H}^{[11]}{\bf v}^{[21]}_ix^{[21]}_i+\sum_{i=1}^{d_{12}}{\bf H}^{[12]}{\bf v}^{[12]}_ix^{[12]}_i+\sum_{i=1}^{d_{22}}{\bf H}^{[12]}{\bf v}^{[22]}_ix^{[22]}_i+{\bf N}^{[1]}.
\end{eqnarray*}

We wish to calculate the dimensionality of the range space of the interference.
There can be three kinds of terms in the interference. The first are those that are zero forced by the transmitter. Second, there are pairs of interference vectors that are aligned (linearly dependent) so that each pair only spans one dimension as explained in (\ref{eq:shadow}). The remaining terms contribute one dimension each. 

Mathematically, the interference signal is expressed as
\begin{eqnarray*}
&&\sum_{i=1}^{d_{21}}{\bf H}^{[11]}{\bf v}^{[21]}_ix^{[21]}_i+\sum_{i=1}^{d_{22}}{\bf H}^{[12]}{\bf v}^{[22]}_ix^{[22]}_i\\
&=&\overbrace{\sum_{i=1}^{\min(d_{21},r_1)}{\bf H}^{[11]}{\bf v}^{[21]}_ix^{[21]}_i+\sum_{i=1}^{\min(d_{22},r_2)}{\bf H}^{[12]}{\bf v}^{[22]}_ix^{[22]}_i}^{=0}+
\underbrace{\sum_{i=1}^{r_0}\left(
\overbrace{{\bf H}^{[11]}{\bf v}^{[21]}_{r_1+i}x^{[21]}_{r_1+i}+{\bf H}^{[12]}{\bf v}^{[22]}_{r_2+i}x^{[22]}_{r_2+i}}^{\mbox{\small range space dimension }~=~1}
\right)}_{\mbox{\small range space dimension }~=~r_0}\\
&&+\underbrace{\sum_{i=r_1+r_0+1}^{d_{21}}{\bf H}^{[11]}{\bf v}^{[21]}_ix^{[21]}_i}_{\mbox{\small range space dimension }~=~(d_{21}-r_1)^+-r_0}
+\underbrace{\sum_{i=r_2+r_0+1}^{d_{22}}{\bf H}^{[12]}{\bf v}^{[22]}_ix^{[22]}_i}_{\mbox{\small range space dimension }~=~(d_{22}-r_2)^+-r_0}.
\end{eqnarray*}
where $r_0$ is the number of pairs of linearly dependent (aligned) interference vectors. Counting dimensions, we obtain:
\begin{eqnarray}
&&\#\mbox{Interference dimensions zero forced by the transmitter}=\min(d_{21},r_1)+\min(d_{22},r_2).\\
&&\#\mbox{Overlapping dimensions}=\min\left[(d_{21}-r_1)^+, (d_{22}-r_2)^+,r\right]\define r_0.\\
&&\#\mbox{Non-overlapping interference dimensions (due to $W_{21})$}=(d_{21}-r_1)^+-r_0.\\
&&\#\mbox{Non-overlapping interference dimensions (due to $W_{22})$}=(d_{22}-r_2)^+-r_0.\\
&&\mbox{Total number of interference dimensions}= (d_{21}-r_1)^++(d_{22}-r_2)^+-r_0.
\end{eqnarray}

Substituting the total number of interference dimensions into the condition (\ref{eq:condition}), and switching indices 1 and 2 to obtain the corresponding condition for receiver 1, we conclude that $(d_{11}, d_{12}, d_{21}, d_{22})$ is achievable provided:
\begin{eqnarray}
N_1&\geq& d_{11}+d_{12}+(d_{21}-(M_1-N_1)^+)^++(d_{22}-(M_2-N_1)^+)^+\nonumber\\
&&~-\min\left[\scriptstyle{(d_{21}-(M_1-N_1)^+)^+, (d_{22}-(M_2-N_1)^+)^+,(M_1+M_2-N_1)^+-(M_1-N_1)^+-(M_2-N_1)^+}\right].\label{eq:achcond1}\\
N_2&\geq& d_{22}+d_{21}+(d_{12}-(M_2-N_2)^+)^++(d_{11}-(M_1-N_2)^+)^+\nonumber\\
&&~-\min\left[\scriptstyle{(d_{12}-(M_2-N_2)^+)^+, (d_{11}-(M_1-N_2)^+)^+,(M_1+M_2-N_2)^+-(M_1-N_2)^+-(M_2-N_2)^+}\right].\label{eq:achcond2}
\end{eqnarray}
Next we show that all  $(d_{11}, d_{12}, d_{21}, d_{22})$ in $\mathcal{D}^Z_{in}$ satisfy both conditions. Starting with condition (\ref{eq:achcond1})

Case 1: $\min\left[\scriptstyle{(d_{21}-(M_1-N_1)^+)^+, (d_{22}-(M_2-N_1)^+)^+,(M_1+M_2-N_1)^+-(M_1-N_1)^+-(M_2-N_1)^+}\right]=(d_{21}-(M_1-N_1)^+)^+$:
\begin{eqnarray}
\mbox{Condition} (\ref{eq:achcond1})\Leftrightarrow N_1\geq d_{11}+d_{12}+d_{22}-(M_2-N_1)^+\Leftrightarrow \max(N_1,M_2)\geq d_{11}+d_{12}+d_{22}
\end{eqnarray}

Case 2: $\min\left[\scriptstyle{(d_{21}-(M_1-N_1)^+)^+, (d_{22}-(M_2-N_1)^+)^+,(M_1+M_2-N_1)^+-(M_1-N_1)^+-(M_2-N_1)^+}\right]=(d_{22}-(M_2-N_1)^+)^+$: 
\begin{eqnarray}
\mbox{Condition} (\ref{eq:achcond1})\Leftrightarrow N_1\geq d_{11}+d_{12}+d_{21}-(M_1-N_1)^+\Leftrightarrow \max(N_1,M_1)\geq d_{11}+d_{12}+d_{21}
\end{eqnarray}

Case 3: $\min\left[\scriptstyle{(d_{21}-(M_1-N_1)^+)^+, (d_{22}-(M_2-N_1)^+)^+,(M_1+M_2-N_1)^+-(M_1-N_1)^+-(M_2-N_1)^+}\right]=(M_1+M_2-N_1)^+-(M_1-N_1)^+-(M_2-N_1)^+$
\begin{eqnarray}
\mbox{Condition} (\ref{eq:achcond1})\Leftrightarrow N_1\geq d_{11}+d_{12}+d_{21}+d_{22}-(M_1+M_2-N_1)^+\nonumber\\
\Leftrightarrow \max(N_1,M_1+M_2)\geq d_{11}+d_{12}+d_{21}+d_{22}
\end{eqnarray}
Thus, in each case we end up with a condition that applies to all $(d_{11}, d_{12}, d_{21}, d_{22})$ in $\mathcal{D}^X_{in}$. It can be similarly verified that Condition (\ref{eq:achcond2}) holds for all $(d_{11}, d_{12}, d_{21}, d_{22})$ in $\mathcal{D}^X_{in}$. Thus, we conclude that all $(d_{11}, d_{12}, d_{21}, d_{22})$ in $\mathcal{D}^X_{in}$ are achievable. By time-sharing their convex hull is achievable as well and the achievability proof is complete.
\end{proof}

\section{Total degrees of freedom on the MIMO $X$ channel}
While the set $\mathcal{D}^X_{out}$ provides an outerbound for all achievable $d_{ij}$ on the MIMO $X$ channel, maximizing any weighted sum of $d_{ij}$ over $\mathcal{D}^X_{out}$ is a linear programming problem. The following theorem presents an outerbound $\eta_{out}$ for the total degrees of freedom $\eta_X^\star$ in closed form by explicitly solving the linear programming problem.
\begin{theorem}\label{theorem:sumdr}
\begin{eqnarray}
\eta_X^\star~\leq~\eta_{out}&\define&\max_{\mathcal{D}^X_{out}}(d_{11}+d_{12}+d_{21}+d_{22})\nonumber\\
&=&\min\left\{
\begin{array}{c}
M_1+M_2, N_1+N_2,\\
\frac{\max(M_1,N_1)+\max(M_1,N_2)+M_2}{2},\\
\frac{\max(M_2,N_1)+\max(M_2,N_2)+M_1}{2}, \\
\frac{\max(M_1,N_1)+\max(M_2,N_1)+N_2}{2}, \\
\frac{\max(M_1,N_2)+\max(M_2,N_2)+N_1}{2}, \\
\frac{\max(M_1,N_1)+\max(M_1,N_2)+\max(M_2,N_1)+\max(M_2,N_2)}{3}
\end{array}
\right\}
\end{eqnarray}
\end{theorem}
\begin{proof}
The theorem is proved by solving the dual problem for the linear program $\max_{\mathcal{D}^X_{out}}(d_{11}+d_{12}+d_{21}+d_{22})$. We explicitly evaluate all the extreme points of the feasible space, calculate the objective value at the extreme points and eliminate the redundant bounds.  Using the fundamental theorem of linear programming we have the result of Theorem \ref{theorem:sumdr}. The details of the derivation are omitted for brevity.
\end{proof}
Note that all 7 terms in the min expression of Theorem \ref{theorem:sumdr} are necessary in general. The following examples illustrate this point, as in each case only one of the $7$ bounds is tight.
\begin{eqnarray*}
\mbox{Example 1:}& M_1=1, M_2=1, N_1=2, N_2=2& \Rightarrow \eta_{out}=M_1+M_2=2\\
\mbox{Example 2:}& M_1=4, M_2=8, N_1=6, N_2=10& \Rightarrow \eta_{out}=\frac{\max(M_2,N_1)+\max(M_2,N_2)+M_1}{2}=11 \nonumber\\
\mbox{Example 3:}& M_1=3, M_2=3, N_1=3, N_2=3& \Rightarrow \eta_{out}=\frac{\scriptstyle{\max(M_1,N_1)+\max(M_1,N_2)+\max(M_2,N_1)+\max(M_2,N_2)}}{3}=4.
\end{eqnarray*}

Similarly, in order to calculate the corresponding lowerbound for $\eta_X^\star$ we need to compute $\max_{\mathcal{D}_{in}^X}(d_{11}+d_{12}+d_{21}+d_{22})$. However, we do not pursue this path due to the well known complexity of integer linear programming. Instead, we show through simple arguments that the total degrees of freedom on the $X$ channel can exceed those achievable by its constituent multiple access, broadcast and interference channels by at most a factor of $\frac{4}{3}$.
\subsection{MAC-BC-IC Innerbound}
Since the MIMO MAC, BC and interference channels are contained in the MIMO $X$ channel and the maximum multiplexing gain for each of these channels is known, we can identify the following MAC-BC-IC innerbound $\eta_{MBI}$ on the maximum multiplexing gain of the MIMO $X$ channel.
\begin{eqnarray*}
\eta_{MBI}&\geq& \min(M_1+M_2, N_1)\\
\eta_{MBI}&\geq& \min(M_1+M_2, N_2)\\
\eta_{MBI}&\geq& \min(M_1, N_1+N_2)\\
\eta_{MBI}&\geq& \min(M_2, N_1+N_2)\\
\eta_{MBI}&\geq& \min(M_1+M_2, N_1+N_2, \max(M_1,N_2),\max(M_2,N_1))\\
\eta_{MBI}&\geq& \min(M_1+M_2, N_1+N_2, \max(M_1,N_1),\max(M_2,N_2))
\end{eqnarray*}
The first two inequalities represent the achievable multiplexing gain for the multiple access channels contained in the $X$ channel. The second set of two inequalities represent the achievable multiplexing gain for the broadcast channels contained in the $X$ channel. The last set of two inequalities represent the achievable multiplexing gain for the interference channels contained in the $X$ channel. The union of these innerbounds can be collectively defined as the MAC-BC-IC innerbound  $\eta_{MBI}$ and can be simplified into the following form:
\begin{eqnarray}
\eta_{MBI}&=& \min(M_1+M_2, N_1+N_2, \max(M_1,N_1,M_2,N_2))
\end{eqnarray}
The following theorem narrows the gap between the innerbound and the outerbound on the multiplexing gain for the MIMO $X$ channel.
\begin{theorem}\label{theorem:zfx}
The maximum total degrees of freedom for the MIMO $X$ channel cannot be more than $4/3$ times the MAC-BC-IC innerbound:
\begin{eqnarray}
 \eta_X^\star\leq\frac{4}{3}\eta_{MBI}
\end{eqnarray}
\end{theorem}
\begin{proof}
The proof is straightforward since,
\begin{eqnarray*}
\frac{4}{3}\max(M_1,N_1,M_2,N_2) \geq \frac{\max(M_1,N_1)+\max(M_1,N_2)+\max(M_2,N_1)+\max(M_2,N_2)}{3}).
\end{eqnarray*}
\end{proof}

\begin{corollary}\label{cor:zfx}
\begin{eqnarray*}
\frac{4}{3}\eta_{MBI} \geq  \eta_X,
\end{eqnarray*}
\end{corollary}
\begin{proof}
Since the RHS is an upperbound on $\eta_X$, the proof of Corollary \ref{cor:zfx} follows directly from the statement of Theorem \ref{theorem:zfx}. 
\end{proof}

Therefore, we have established that zero forcing techniques can achieve at least three fourths of the maximum multiplexing gain of the MIMO $X$ channel. An interesting case is when $M_1=M_2=N_1=N_2=M$ for which the MMK scheme can achieve the maximum possible multiplexing gain $\frac{4}{3}M$ rounded down to the nearest integer, i.e. $\lfloor\frac{4}{3}M\rfloor$.

\section{Equal antennas at all nodes $M_1=M_2=N_1=N_2=M$}
\subsection{Zero Forcing and the MMK Scheme}
The MMK (Maddeh-Ali-Motahari-Khandani) scheme is an elegant coding scheme for the MIMO $X$ channel. In \cite{MMK} it is shown that for a MIMO $X$ channel with $M=3$ antennas at all nodes, a multiplexing gain of $4$ is achievable by a combination of dirty paper coding, successive decoding and zero forcing. For general $M$, the MMK scheme achieves $\lfloor\frac{4}{3}M\rfloor$ degrees of freedom. 

Comparing these results to our inner and outerbounds, note that it is straightforward to obtain $\eta_X^\star\leq\frac{4}{3}M$ by substituting $M_1=M_2=N_1=N_2=M$ into the upperbound of Theorem \ref{theorem:sumdr}. 

Interestingly, our zero forcing based scheme described in the proof of Theorem \ref{theorem:innerbound} also suffices to achieve the $\lfloor\frac{4}{3}M\rfloor$ degrees of freedom. The zero forcing achievability result is verified as follows. We write $M=3m+k$ where $k$ is either $0,1$ or $2$. Therefore $\lfloor\frac{4}{3}M\rfloor=4m+k$. Let us assign:
\begin{eqnarray}
d_{11}=m+k\\
d_{12}=d_{21}=d_{22}=m
\end{eqnarray}
With these values it is easy to verify that $(d_{11},d_{12},d_{21},d_{22})\in\mathcal{D}_{in}^X$. Thus, $\lfloor\frac{4}{3}M\rfloor=4m+k$ degrees of freedom are achievable with zero forcing.

Thus we have established that for the MIMO $X$ channel with $M$ antennas at all nodes the degrees of freedom for the sum rate are bounded as $\lfloor\frac{4}{3}M\rfloor\leq\eta_X^\star\leq\frac{4}{3}M$. In other words we are always able to achieve the maximum degrees of freedom rounded down to the nearest integer. This is optimal when $M$ is a multiple of $3$. However, when $M$ is not a multiple of $3$ the inner and outerbounds differ by a fraction equal to either $1/3$ or $2/3$. The achievability of the remaining fractional degrees of freedom is an issue that touches upon a fundamental question of the optimality of fractional degrees of freedom. In general it is not known whether there are wireless networks with non integer valued degrees of freedom. The following theorem shows that indeed the MIMO $X$ channel can have non-integer valued degrees of freedom and also establishes the precise degrees of freedom for the MIMO $X$ channel with $M>1$ antennas at all nodes.
\subsection{The Degrees of Freedom for $M>1$}
Throughout this paper we assume channel matrices are generated from a continuous distribution and therefore the channels are non-degenerate with probability one so that the degrees of freedom characterizations are valid almost surely. However, the following theorem is an exception as it applies to any \emph{arbitrary} channel matrices that satisfy the conditions specified in the theorem. The result of Theorem \ref{theorem:noninteger} is stronger than we need for Corollary \ref{corollary:noninteger} which is the main result of this section. However, we prove the stronger result in Theorem \ref{theorem:noninteger} because it will be useful when we discuss time/frequency varying channel coefficients in Section \ref{sec:timevarying}.
\begin{theorem}\label{theorem:noninteger}
For the MIMO $X$ channel with $M>1$ antennas at all nodes and \emph{arbitrary} $M\times M$ channel matrices ${\bf H}^{[11]}, {\bf H}^{[12]},{\bf H}^{[21]},{\bf H}^{[22]}$, the degrees of freedom 
\begin{equation}
\eta^\star = \frac{4}{3}M
\end{equation}
if the channel matrices ${\bf H}^{[11]}, {\bf H}^{[12]},{\bf H}^{[21]},{\bf H}^{[22]}$  are invertible and the product channel matrix:
\begin{equation}
{\bf F}\define\left( {\bf H}^{[11]}\right)^{-1} {\bf H}^{[12]}\left( {\bf H}^{[22]}\right)^{-1}{\bf H}^{[21]}\label{eq:defineF}
\end{equation}
is nondefective and ${\bf F}$ does not have any eigenvalues with multiplicity more than $\frac{2}{3}M$.
\end{theorem}

\begin{corollary}\label{corollary:noninteger}
For the MIMO $X$ channel with $M>1$ antennas at all nodes with channel matrices  generated from a continuous distribution
\begin{equation}
\eta^\star = \frac{4}{3}M
\end{equation}
with probability one.
\end{corollary}
Corollary \ref{corollary:noninteger} follows directly from the result of Theorem \ref{theorem:noninteger} because when the channel coefficients are generated from a continuous distribution then all matrices are non-singular and all eigenvalues of ${\bf F}$ are distinct with probability one, so that all the conditions of Theorem \ref{theorem:noninteger} are satisfied almost surely.

Before proving the result of Theorem \ref{theorem:noninteger} we explain the significance of the conditions required by the statement of the theorem. First, we note that the non-singularity condition required by Theorem \ref{theorem:noninteger} is not always necessary for achievability of $4M/3$ degrees of freedom. For example, consider the case $M_1=2,M_2=2,N_1=3,N_2=3$ which can be considered to be a special case of the $M_1=M_2=N_1=N_2=3$ scenario with rank deficient channel matrices, i.e. each transmitter has a third antenna with zero channel gain. The achievable degrees of freedom region established in Theorem \ref{theorem:innerbound}  shows that $d_{11}=d_{12}=d_{21}=d_{22}=1$ is achievable for this MIMO $X$ channel so that the total number of degrees of freedom is $4$ (the upperbound is tight) even with all $3\times 3$ channel matrices rank deficient (each with rank $2$). 

Second, the constraint on the multiplicity of the eigenvalues of ${\bf F}$ is interesting and may be necessary. For $M=1$ note that the multiplicity constraint cannot be satisfied because the multiplicity of an eigenvalue cannot be less than 1. For $M=2$ we need the two eigenvalues of ${\bf F}$ to be distinct, otherwise the $M=2$ case is identical to the $M=1$ case. For $M=3$ it is sufficient to have at least 1 distinct eigenvalue. For $M=4,5,6, \cdots$ we require that all the eigenvalues should have multiplicities less than $4, 4, 5, \cdots$, respectively. Note that with the exception of $M=1$ case all these conditions are true with probability one if the channels are generated according to a continuous probability distribution.

The key to the proof is to consider a $3$ symbol extension of the channel so that we have effectively a $3M\times 3M$ channel, over which we will achieve $4M$ degrees of freedom. Note that we still assume the channel matrices are fixed, so that the $3$ symbol extension does not provide us a new channel matrix over each slot. Instead, each $M\times M$ channel matrix is repeated three times to produce a $3M\times 3M$ block diagonal matrix. 

The three symbol extension of the channel does not trivially solve the problem, as is evident from the $M=1$ case. For $M=1$, the 3-symbol extension channel gives us:

\begin{eqnarray}
\overline{\bf Y}^{[1]}&=&\overline{\bf H}^{[11]}\overline{\bf X}^{[1]}+\overline{\bf H}^{[12]}\overline{\bf X}^{[2]}+\overline{\bf N}^{[1]}\\
\overline{\bf Y}^{[2]}&=&\overline{\bf H}^{[21]}\overline{\bf X}^{[1]}+\overline{\bf H}^{[22]}\overline{\bf X}^{[2]}+\overline{\bf N}^{[2]}
\end{eqnarray}
where the overbar notation represents the 3-symbol extensions so that
\begin{eqnarray*}
\overline{\bf A}\define\left[\begin{array}{c}{\bf A}(3n)\\{\bf A}(3n+1)\\{\bf A}(3n+2)\end{array}\right], 
\end{eqnarray*}
when ${\bf A}$ is a $M\times 1$ vector that takes value ${\bf A}(n)$ at time $n$, and  
\begin{eqnarray*}
\overline{\bf H}\define\left[\begin{array}{ccc}{\bf H}&{\bf 0}&{\bf 0}\\{\bf 0}&{\bf H}&{\bf 0}\\{\bf 0}&{\bf 0}&{\bf H}\end{array}\right], 
\end{eqnarray*}
when ${\bf H}$ is an $M\times M$ channel matrix. For the $M=1$ case the channel matrix $\overline{\bf H}^{[ij]}=H^{[ij]}{\bf I}$ is simply a scalar multiple of the identity matrix, so that
\begin{eqnarray}
\overline{\bf Y}^{[1]}&=&{H}^{[11]}\overline{\bf X}^{[1]}+H^{[12]}\overline{\bf X}^{[2]}+\overline{\bf N}^{[1]}\\
\overline{\bf Y}^{[2]}&=&H^{[21]}\overline{\bf X}^{[1]}+H^{[22]}\overline{\bf X}^{[2]}+\overline{\bf N}^{[2]}
\end{eqnarray}
Thus, the alignment of the received signal vectors $\overline{\bf X}^{[1]}, \overline{\bf X}^{[2]}$ is identical at both receivers. This makes it impossible to have the spatial directions of the signals align at one receiver (where they are treated as interference) and remain distinct at the other receiver (where they represent the desired signals). The apparent problem with the 3-symbol extension model for the $M=1$ case makes it surprising that the same idea works for $M>1$. The details of the proof for $M>1$ are presented below.

\begin{proof} 
We assign equal number of degrees of freedom $d_{11}=d_{12}=d_{21}=d_{22}=M$ to all four messages for a total of $4M$ degrees of freedom over the 3-symbol extension channel defined above.
Transmitter $j$ sends message $W_{ij}$ to receiver $i$ in the form of $M$ independently encoded streams along the direction vectors $\overline{\bf v}^{[ij]}_1, \overline{\bf v}^{[ij]}_2, \cdots, \overline{\bf v}^{[ij]}_M$, each of dimension $3M\times 1$, so that we have:
\begin{eqnarray}
\overline{\bf X}^{[1]}&=&\sum_{m=1}^M\overline{\bf v}^{[11]}_mx^{[11]}_m+\sum_{m=1}^M\overline{\bf v}^{[21]}_mx^{[21]}_m=\overline{\bf V}^{[11]}{\bf X}^{[11]}+\overline{\bf V}^{[21]}{\bf X}^{[21]}\\
\overline{\bf X}^{[2]}&=&\sum_{m=1}^M\overline{\bf v}^{[12]}_mx^{[12]}_m+\sum_{m=1}^M\overline{\bf v}^{[22]}_mx^{[22]}_m=\overline{\bf V}^{[12]}{\bf X}^{[12]}+\overline{\bf V}^{[22]}{\bf X}^{[22]}
\end{eqnarray}
where the ${\bf V}^{[ij]}$ and ${\bf X}^{[ij]}$ are $3M\times M$ and $M\times 1$ matrices respectively.
Interference alignment is achieved by setting
\begin{eqnarray}
\overline{\bf H}^{[11]}\overline{\bf V}^{[21]}=\overline{\bf H}^{[12]}\overline{\bf V}^{[22]} &\Leftrightarrow& \overline{\bf V}^{[22]} = \left(\overline{\bf H}^{[12]}\right)^{-1}\overline{\bf H}^{[11]}\overline{\bf V}^{[21]}\label{eq:transmitdirection1}\\
\overline{\bf H}^{[21]}\overline{\bf V}^{[11]}=\overline{\bf H}^{[22]}\overline{\bf V}^{[12]} &\Leftrightarrow& \overline{\bf V}^{[12]} = \left(\overline{\bf H}^{[22]}\right)^{-1}\overline{\bf H}^{[21]}\overline{\bf V}^{[11]}\label{eq:transmitdirection2}
\end{eqnarray}
So once we pick the direction vectors $\overline{\bf V}^{[11]},\overline{\bf V}^{[21]}$ for transmitter $1$ the direction vectors $\overline{\bf V}^{[12]},\overline{\bf V}^{[22]}$ for transmitter $2$ are automatically determined.
With these choices, the output signals at receivers 1 and 2 are the $3M\times 1$ vectors,
\begin{eqnarray}
\overline{\bf Y}^{[1]}&=&\overline{\bf H}^{[11]}\overline{\bf V}^{[11]}{\bf X}^{[11]}+\overline{\bf H}^{[11]}\overline{\bf V}^{[21]}{\bf X}^{[21]}+\overline{\bf H}^{[12]}\overline{\bf V}^{[12]}{\bf X}^{[12]}+\overline{\bf H}^{[12]}\overline{\bf V}^{[22]}{\bf X}^{[22]}+\overline{\bf N}^{[1]}\\
&=&\overline{\bf H}^{[11]}\overline{\bf V}^{[11]}{\bf X}^{[11]}+\overline{\bf H}^{[12]}\left(\overline{\bf H}^{[22]}\right)^{-1}\overline{\bf H}^{[21]}\overline{\bf V}^{[11]}{\bf X}^{[12]}+\overline{\bf H}^{[11]}\overline{\bf V}^{[21]}\left({\bf X}^{[21]}+{\bf X}^{[22]}\right)+\overline{\bf N}^{[1]}\\
\overline{\bf Y}^{[2]}&=&\overline{\bf H}^{[21]}\overline{\bf V}^{[11]}{\bf X}^{[11]}+\overline{\bf H}^{[21]}\overline{\bf V}^{[21]}{\bf X}^{[21]}+\overline{\bf H}^{[22]}\overline{\bf V}^{[12]}{\bf X}^{[12]}+\overline{\bf H}^{[22]}\overline{\bf V}^{[22]}{\bf X}^{[22]}+\overline{\bf N}^{[2]}\\
&=&\overline{\bf H}^{[21]}\overline{\bf V}^{[21]}{\bf X}^{[21]}+\overline{\bf H}^{[22]} \left(\overline{\bf H}^{[12]}\right)^{-1}\overline{\bf H}^{[11]}\overline{\bf V}^{[21]}{\bf X}^{[22]}+\overline{\bf H}^{[21]}\overline{\bf V}^{[11]}\left({\bf X}^{[11]}+{\bf X}^{[12]}\right)+\overline{\bf N}^{[2]}
\end{eqnarray}
With the interfering signals already aligned, each receiver can separate the signal and interference signals provided all the received directions are linearly independent. In other words, we must pick ${\bf V}^{[11]}, {\bf V}^{[21]}$ such that each of the two $3M\times 3M$matrices:
\begin{eqnarray*}
\left[\overline{\bf H}^{[11]}\overline{\bf V}^{[11]}  ~~~ \overline{\bf H}^{[12]}\left(\overline{\bf H}^{[22]}\right)^{-1}\overline{\bf H}^{[21]}\overline{\bf V}^{[11]} ~~~~ \overline{\bf H}^{[11]}\overline{\bf V}^{[21]}\right],~~~ \left[\overline{\bf H}^{[21]}\overline{\bf V}^{[21]}~~~ \overline{\bf H}^{[22]} \left(\overline{\bf H}^{[12]}\right)^{-1}\overline{\bf H}^{[11]}\overline{\bf V}^{[21]} ~~~~ \overline{\bf H}^{[21]}\overline{\bf V}^{[11]} \right] 
\end{eqnarray*}
has full rank $3M$. Since multiplication with a nonsingular matrix does not affect the rank of a matrix, we require (equivalently) that each of the two $3M\times 3M$matrices:
\begin{eqnarray*}
\left[\overline{\bf V}^{[11]}  ~~~\overline{\bf F}\overline{\bf V}^{[11]} ~~~~\overline{\bf V}^{[21]}\right],~~~ \left[\overline{\bf V}^{[21]}~~~  \overline{\bf F}^{-1}\overline{\bf V}^{[21]} ~~~~ \overline{\bf V}^{[11]} \right] 
\end{eqnarray*}
must have full rank $3M$, where ${\bf F} = \left( {\bf H}^{[11]}\right)^{-1} {\bf H}^{[12]}\left( {\bf H}^{[22]}\right)^{-1}{\bf H}^{[21]}$ is the $M\times M$ matrix defined in (\ref{eq:defineF}) and $\overline{F}$ is the 3-symbol extension of ${\bf F}$ into a $3M\times 3M$ block diagonal matrix.
Since ${\bf F}$ has $M$ non-zero and distinct eigenvalues by the assumptions of Theorem \ref{theorem:noninteger}, it follows that $\overline{F}$ has $3M$ linearly independent eigenvectors and these eigenvectors form a complete basis (not necessarily an orthogonal basis) for the entire $3M$ dimensional vector space. For simplicity we wish to align our coordinate system to the eigenbasis of $\overline{\bf F}$. To this end, let $\overline{\bf E}_F$ be the $3M\times 3M$ matrix whose column vectors are the eigenvectors of $\overline{\bf F}$ and let $\overline{\bf \Lambda}=\mbox{Diag}[\overline{\lambda}_1,\cdots,\overline{\lambda}_{3M}]$ be the $3M\times 3M$ matrix containing the eigenvalues of $\overline{\bf F}$ along the main diagonal and zeros elsewhere. 
Without loss of generality we can assume that the eigenvalues of $\overline{\bf F}$ are arranged so that $\overline\lambda_1\neq\overline\lambda_2$, $\overline\lambda_1\neq\overline\lambda_3$, $\overline\lambda_4\neq\overline\lambda_5$, $\overline\lambda_4\neq\overline\lambda_6$, $\cdots$,$\overline\lambda_{3M-2}\neq\overline\lambda_{3M-1}$, $\overline\lambda_{3M-2}\neq\overline\lambda_{3M}$. We explain why there is no loss of generality in this assumption as follows. Note that because we assume that ${\bf F}$ does not have any eigenvalue with multiplicity higher than $2M/3$, it follows that $\overline{\bf F}$ does not have any eigenvalue with multiplicity higher than $2M$. It is always possible to arrange $3M$ eigenvalues into groups of $3$ such that no group consists of all $3$ equal eigenvalues, if and only if there is no repeated eigenvalue with multiplicity higher than $2M$. Since we assume no eigenvalue has multiplicity higher than $2M$ we can arrange the eigenvalues into groups of $3$  as $(\overline\lambda_1,\overline\lambda_2,\overline\lambda_3),(\overline\lambda_4,\overline\lambda_5,\overline\lambda_6),\cdots,(\overline\lambda_{3M-2},\overline\lambda_{3M-1},\overline\lambda_{3M})$ such that the first eigenvalue within each group is distinct from the other two which may or may not be equal.
Note that $\overline{\bf F}$ and $\overline{\bf F}^{-1}$ have the same eigenbasis, so that
\begin{eqnarray}
\overline{\bf F}\overline{\bf E}_F =  \overline{\bf E}_F\overline\Lambda \\
\overline{\bf F}^{-1}\overline{\bf E}_F = \overline{\bf E}_F \overline \Lambda^{-1}\label{eq:eigenbasis}
\end{eqnarray}
Let us define:
\begin{eqnarray}
{\bf V}^{[11]}_F&\define&\overline{\bf E}_F^{-1}\overline{\bf V}^{[11]}~~~\mbox{ so that } ~~~\overline{\bf V}^{[11]}=\overline{\bf E}_F{\bf V}^{[11]}_F\\
{\bf V}^{[21]}_F&\define&\overline{\bf E}_F^{-1}\overline{\bf V}^{[21]}~~~\mbox{ so that } ~~~\overline{\bf V}^{[21]}=\overline{\bf E}_F{\bf V}^{[21]}_F\label{eq:aligntoeigenbasis}
\end{eqnarray}
\begin{eqnarray*}
\mbox{Then, we have rank}\left(\left[\overline{\bf V}^{[11]}  ~~~\overline{\bf F}\overline{\bf V}^{[11]} ~~~~\overline{\bf V}^{[21]}\right]\right)&=&\mbox{rank}\left(\left[\overline{\bf E}_F{\bf V}^{[11]}_F  ~~~\overline{\bf F}\overline{\bf E}_F{\bf V}^{[11]}_F ~~~~\overline{\bf E}_F{\bf V}^{[21]}\right]\right)\\
&=&\mbox{rank}\left(\left[\overline{\bf E}_F{\bf V}^{[11]}_F  ~~~ \overline{\bf E}_F\overline\Lambda{\bf V}^{[11]}_F ~~~~\overline{\bf E}_F{\bf V}^{[21]}_F\right]\right)\\
&=&\mbox{rank}\left(\left[{\bf V}^{[11]}_F  ~~~\overline\Lambda{\bf V}^{[11]}_F ~~~~{\bf V}^{[21]}_F\right]\right)\\
\mbox{and similarly, rank}\left(\left[\overline{\bf V}^{[21]}  ~~~\overline{\bf F}^{-1}\overline{\bf V}^{[21]} ~~~~\overline{\bf V}^{[11]}\right]\right)&=&\mbox{rank}\left(\left[{\bf V}^{[21]}_F  ~~~\overline\Lambda^{-1}{\bf V}^{[21]}_F ~~~~{\bf V}^{[11]}_F\right]\right)\\
\end{eqnarray*}
We need to prove that we can pick ${\bf V}^{[11]}_F, {\bf V}^{[21]}_F$ so that \[\mbox{rank}\left(\left[{\bf V}^{[21]}_F  ~~~\overline\Lambda^{-1}{\bf V}^{[21]}_F ~~~~{\bf V}^{[11]}_F\right]\right)=\mbox{rank}\left(\left[{\bf V}^{[11]}_F  ~~~\overline\Lambda{\bf V}^{[11]}_F ~~~~{\bf V}^{[21]}_F\right]\right)=3M\]
This is proven by the following explicit choice of the $M\times 3M$ matrices ${\bf V}^{[11]}_F, {\bf V}^{[21]}_F$:
\begin{eqnarray*}
{\bf V}^{[11]}_F=\left[\begin{array}{cccc}\boxed{\begin{array}{c}1\\1\\0\end{array}}&{\bf 0}_{3\times 1}&\cdots&{\bf 0}_{3\times 1}\\
{\bf 0}_{3\times 1}&\boxed{\begin{array}{c}1\\1\\0\end{array}}&\cdots&{\bf 0}_{3\times 1}\\
\vdots&\vdots&\ddots&\vdots\\
{\bf 0}_{3\times 1}&{\bf 0}_{3\times 1}&\cdots&\boxed{\begin{array}{c}1\\1\\0\end{array}}
\end{array}\right],
{\bf V}^{[21]}_F=\left[\begin{array}{cccc}\boxed{\begin{array}{c}1\\0\\1\end{array}}&{\bf 0}_{3\times 1}&\cdots&{\bf 0}_{3\times 1}\\
{\bf 0}_{3\times 1}&\boxed{\begin{array}{c}1\\0\\1\end{array}}&\cdots&{\bf 0}_{3\times 1}\\
\vdots&\vdots&\ddots&\vdots\\
{\bf 0}_{3\times 1}&{\bf 0}_{3\times 1}&\cdots&\boxed{\begin{array}{c}1\\0\\1\end{array}}
\end{array}\right]
\end{eqnarray*}
Thus ${\bf V}^{[11]}_F$ is formed from the block diagonal repetitions of the $[1~~~1~~~0]^T$ column vector while ${\bf V}^{[21]}_F$ is formed from the block diagonal repetitions of the $[1~~~0~~~1]^T$ column vector.
With this choice of ${\bf V}^{[11]}_F, {\bf V}^{[21]}_F$ we now show that:
\begin{eqnarray*}
\left[{\bf V}^{[11]}_F  ~~~\overline\Lambda{\bf V}^{[11]}_F ~~~~{\bf V}^{[21]}_F\right]{\bf A}={\bf 0} &\mbox{only if}& {\bf A}={\bf 0} \mbox{ which means rank}\left[{\bf V}^{[11]}_F  ~~~\overline\Lambda{\bf V}^{[11]}_F ~~~~{\bf V}^{[21]}_F\right]=3M\\
\left[{\bf V}^{[21]}_F  ~~~\overline\Lambda^{-1}{\bf V}^{[21]}_F ~~~~{\bf V}^{[11]}_F\right]{\bf B}={\bf 0} &\mbox{only if}& {\bf B}={\bf 0}\mbox{ which means rank}\left[{\bf V}^{[21]}_F  ~~~\overline\Lambda^{-1}{\bf V}^{[21]}_F ~~~~{\bf V}^{[11]}_F\right]=3M
\end{eqnarray*}
for any $3M\times 1$ vectors ${\bf A}, {\bf B}$. 

Suppose $\left[{\bf V}^{[11]}_F  ~~~\overline\Lambda{\bf V}^{[11]}_F ~~~~{\bf V}^{[21]}_F\right]{\bf A}={\bf 0}$. The case for $M=3$ is illustrated explicitly below as an example while the analysis is for general $M>1$.
\begin{eqnarray*}
\left[\begin{array}{ccccccccc}
1&0&0& \overline\lambda_1&0&0&  1&0&0\\
1&0&0& \overline\lambda_2&0&0&  0&0&0\\
0&0&0& 0&0&0&  1&0&0\\
0&1&0& 0&\overline\lambda_4&0&  0&1&0\\
0&1&0& 0&\overline\lambda_5&0&  0&0&0\\
0&0&0& 0&0&0&  0&1&0\\
0&0&1& 0&0&\overline\lambda_7&  0&0&1\\
0&0&1& 0&0&\overline\lambda_8&  0&0&0\\
0&0&0& 0&0&0&  0&0&1
\end{array}\right]
\left[\begin{array}{c}
a_1\\a_2\\a_3\\a_4\\a_5\\a_6\\a_7\\a_8\\a_9
\end{array}\right]
=
\left[\begin{array}{c}
0\\0\\0\\0\\0\\0\\0\\0\\0
\end{array}\right]
\end{eqnarray*}
The equations corresponding to the first $3$ rows:
\begin{eqnarray*}
a_1+\overline\lambda_1a_{M+1}&=&-a_{2M+1}\\
a_1+\overline\lambda_2a_{M+1}&=&0\\
a_{2M+1}&=&0
\end{eqnarray*}
Since $\overline\lambda_1\neq\overline\lambda_2$, the simultaneous equations imply $a_{1}=a_{M+1}=a_{2M+1}=0$. Similarly, for each $i\in\{0,1,\cdots,M-1\}$, considering the simultaneous equations corresponding to rows $3*i+1, 3*i+2, 3*i+3$ we have 
\begin{eqnarray}
a_{i+1}+\overline\lambda_{3*i+1}a_{M+i+1}&=&-a_{2M+i+1}\\
a_{i+1}+\overline\lambda_{3*i+2}a_{M+i+1}&=&0\\
a_{2M+i+1}&=&0
\end{eqnarray}
which implies $a_{i+1}=a_{M+i+1}=a_{2M+i+1}=0$. Combining the results, we have ${\bf A}={\bf 0}$, i.e. $\left[{\bf V}^{[11]}_F  ~~~\overline\Lambda{\bf V}^{[11]}_F ~~~~{\bf V}^{[21]}_F\right]{\bf A}$ must have rank $3M$.

Similarly let $\left[{\bf V}^{[21]}_F  ~~~\overline\Lambda^{-1}{\bf V}^{[21]}_F ~~~~{\bf V}^{[11]}_F\right]{\bf B}={\bf 0}$. The first three rows in that case provide us with equations:
\begin{eqnarray}
b_1+\frac{1}{\overline\lambda_1}b_{M+1}&=&-b_{2M+1}\\
b_{2M+1}&=&0\\
b_1+\frac{1}{\overline\lambda_3}b_{M+1}&=&0
\end{eqnarray}
which together imply $b_1=b_{M+1}=b_{2M+1}=0$ because $\lambda_1\neq\lambda_3$. Proceeding as before by considering groups of three rows at a time, we establish that ${\bf B}={\bf 0}$ so that rank$\left[{\bf V}^{[21]}_F  ~~~\overline\Lambda^{-1}{\bf V}^{[21]}_F ~~~~{\bf V}^{[11]}_F\right]=3M$.

Thus all $3M$ received directions ($2M$ for the desired signal and $M$ for the overlapping interference signals) are linearly independent and each receiver can zero force and decode each stream free from interference. Since the $4M$ degrees of freedom are obtained over the $3$-symbol extension channel, the total number of degrees of freedom per channel use is equal to $\frac{4}{3}M$.
\end{proof}

\section{Cooperation through interference alignment and shared messages}
The MIMO $X$ channel and the MIMO interference channel are physically the same channel. They are described by the same input output equations. There is no channel between the transmitters or between the receivers in either case, so no message sharing is possible. However, the $X$ channel is able to achieve more degrees of freedom than the interference channel because of an implicit cooperation in the form of overlapping interference spaces. The key to the higher multiplexing gain of the MIMO $X$ channel is the ability to align signals so that they cast distinct shadows at their desired receivers but overlapping shadows at the unintended receivers. 

What makes this simple idea of interference alignment even more remarkable is that without overlapping interference, i.e. on the interference channel, even explicit cooperation between transmitters and receivers has been found to have no benefit in terms of the degrees of freedom. For the two user interference channel with single antennas at all nodes it is shown in \cite{MadsenIT} and \cite{Nosratinia-Madsen} that even if noisy communication links are introduced between the two transmitters as well as between the two receivers the highest multiplexing gain achievable is equal to one. Message sharing for the interference channel with single antenna nodes has also been studied recently by Devroye and Sharif in \cite{Devroye_Sharif} in the context of cognitive radio. Exploring this idea further, we characterize the degrees of freedom for the MIMO cognitive radio channel, i.e. the MIMO interference channel when some messages are made available in the manner of cognitive radio at either the transmitters or the receivers. This result is also useful for our subsequent development of the MIMO $X$ channel with shared messages. For simplicity we focus primarily on the cases with $M_1=M_2=N_1=N_2=M$.
\subsection{Cognitive MIMO interference channel}
We refer to the interference channel with one message shared between the two transmitters as the \emph{interference channel with a cognitive transmitter}. This channel is often referred to as the cognitive radio channel and is shown in Figure \ref{fig:cogint}. Following cognitive radio terminology, transmitter 1 is the primary transmitter whose message for primary receiver (receiver 1) is known non-causally to transmitter 2, the secondary transmitter. Similarly, an interference channel with a cognitive receiver could be defined as the scenario when the primary user's message is known to the secondary user's receiver. Then there is also the possibility that both the secondary transmitter and secondary receiver have knowledge of the primary user's message. Theorems \ref{theorem:cogint} and \ref{theorem:cogintfull} establish the total degrees of freedom on the MIMO interference channel with cognitive transmitters and/or receivers.
\begin{theorem}\label{theorem:cogint}
On the MIMO interference channel with equal number ($M$) of antennas at all nodes, the total number of degrees of freedom is equal to $M$ for each of the following cognition scenarios (shown in Fig. \ref{fig:cogint}).
\begin{enumerate}
\item  $W_{11}$ is made available non-causally to transmitter $2$ (Fig. \ref{fig:cogint}(a)).
\item $W_{11}$ is made available to receiver $2$  (Fig. \ref{fig:cogint}(b)).
\item $W_{11}$  is made available non-causally to transmitter $2$ and also to receiver $2$ (Fig. \ref{fig:cogint}(c)).
\end{enumerate}
\end{theorem}
\begin{proof} 

The first case mentioned in Theorem \ref{theorem:cogint}, i.e. the cognitive transmitter case, has been shown for $M=1$ in \cite{Devroye_Sharif} while the proof for general $M$ is provided here. We present a proof for the third case above, i.e. with both a cognitive transmitter and a cognitive receiver. Clearly, the first and second cases follow directly from the third case.

Consider any reliable coding scheme for the cognitive interference channel shown in Fig. \ref{fig:cogint}. Since $W_{11}$ is the only message at transmitter 1, its transmitted codeword is completely a function of $W_{11}$. Knowing $W_{11}$ allows receiver $2$ to subtract the contribution of transmitter $1$ from its received signal. For any reliable coding scheme, receiver $1$ by definition must be able to decode $W_{11}$, so it can subtract transmitter 1's signal from its received signal as well. Thus, both receivers are left with a signal coming entirely from transmitter 2. Now, by reducing the noise at receiver 1 as in \cite{Jafar_dof_int} we can argue that if receiver $2$ can decode $W_{22}$ then so can receiver $1$. Thus, receiver $1$ is able to decode both messages and therefore the total degrees of freedom cannot be more than the number of antennas at receiver $1$, which is equal to $M$. Thus, the MIMO interference channel with message $W_{11}$ known to both transmitter $2$ and receiver $2$ cannot have more than $M$ degrees of freedom. The achievability of $M$ degrees of freedom is straightforward.
\end{proof}

Thus, there is no benefit (in terms of total degrees of freedom) from sharing one user's message on the MIMO interference channel with equal number of antennas at all nodes even if this message is shared with both the transmitter and receiver of the other user. However, if the nodes have different number of antennas then the MIMO interference channel may indeed benefit from cognitive message sharing. A simple example is the $(1, n, n, 1)$ case, i.e. transmitter 1 and receiver 2 have one antenna each while receiver 1 and transmitter 2 have $n>1$ antennas. Without message sharing this interference channel has at most $1$ degree of freedom. However, if transmitter 1's message $W_{11}$ is made available non-causally to transmitter 2, then clearly $n$ degrees of freedom can be achieved quite simply by transmitter $2$ sending $W_{11}$ to receiver $1$ on the $n\times n$ channel between them. Message sharing can also be useful for the interference channel even with single antennas at all nodes when channel matrices take certain specialized structured forms as in \cite{Lapidoth_Shamai_Wigger_IN}.

\begin{theorem}\label{theorem:cogintfull}
On the MIMO interference channel with equal number ($M$) of antennas at all nodes, the total number of degrees of freedom is equal to $2M$ for each of the following cognition scenarios (shown in Fig. \ref{fig:cogintfull}).
\begin{enumerate}
\item  $W_{11}$ is made available non-causally to transmitter $2$ and $W_{22}$ is made available non-causally to transmitter $1$.
\item $W_{11}$ is made available to receiver $2$ and $W_{22}$ is made available to receiver $1$.
\item $W_{11}$  is made available non-causally to transmitter $2$ and $W_{22}$ is made available to receiver $1$.
\end{enumerate}
\end{theorem}
\begin{proof}
The converse is trivial because the degrees of freedom cannot exceed the total number of transmit antennas across all users. The achievability arguments for the three cases are as follows. When both transmitters know each other's messages the MIMO interference channel becomes a broadcast channel with a $2M$ antenna transmitter and two receivers with $M$ antennas each and it is well known that this broadcast channel can achieve $2M$ degrees of freedom. The second case, when each receiver knows the message intended for the other receiver is also straightforward as each receiver is able to subtract all the interference and achieve $M$ degrees of freedom for a total of $2M$ degrees of freedom. In the third case, the two transmitters use their $2M$ antennas to zero force the interference seen by the $M$ antennas of receiver $2$ due to the transmission of $W_{11}$. Receiver 1 on the other hand is able to subtract out the interference due to the transmission of the known message $W_{22}$. Thus, neither receiver sees any interference and a total of $2M$ degrees of freedom are achieved.
\end{proof}
Next we explore the degrees of freedom when some messages may be shared between the two transmitters on the MIMO $X$ channel.
\subsection{MIMO $X$ channel with a Cognitive Transmitter}
\begin{theorem}\label{theorem:cogxtransmitter}
On the MIMO $X$ channel with $M>1$ antennas at all nodes if message $W_{11}$ is made available non-causally to transmitter $2$, the total number of degrees of freedom \begin{equation}
\eta^\star = \frac{3}{2}M.
\end{equation}
\end{theorem}
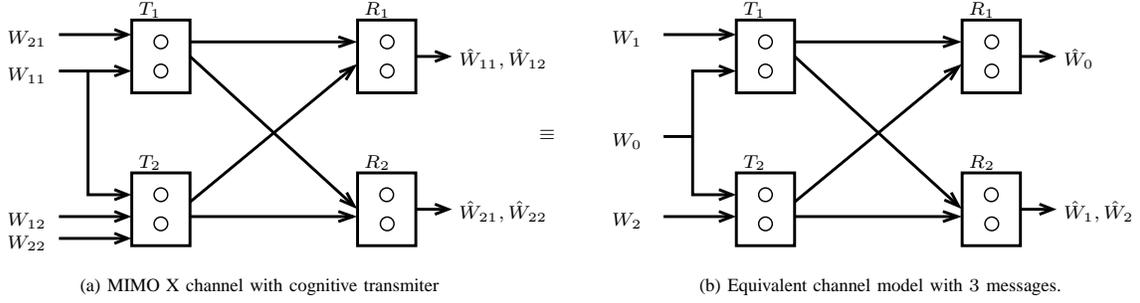
\begin{figure}[h]
\centerline{\input{xscene5.pstex_t}}
\caption{MIMO $X$ channel with a Cognitive Transmitter: $(M>1), \eta = \frac{3}{2}M$}\label{fig:xshared}
\end{figure}




In Fig. \ref{fig:xshared} the message $W_{11}$ is made available to transmitter $2$. Note that regardless of which one of the $4$ messages is made available to the other transmitter, from the total degrees of freedom perspective, all $4$ cases are equivalent to the three message $X$ channel as shown in Fig. \ref{fig:xshared}(b). This channel is called the cognitive $X$ channel in \cite{Devroye_Sharif}.

The argument for the equivalence between the $X$ channel with a cognitive transmitter in Fig. \ref{fig:xshared}(a) and Fig. \ref{fig:xshared}(b)  is as follows. First we show that the total number of degrees of freedom in Fig. \ref{fig:xshared}(b) cannot be more than the $X$ channel with one shared message. This is trivial because  Fig. \ref{fig:xshared}(b) can also be obtained from the $X$ channel with one shared message $W_{11}$ by setting message $W_{12}=\phi, W_{11}=W_0, W_{21}=W_1, W_{22}=W_2$ in the cognitive $X$ channel. 

Next we show that  the total number of degrees of freedom for Fig. \ref{fig:xshared}(b) cannot be smaller than the $X$ channel with only one shared message $W_{11}$. To this end, suppose, in addition to providing $W_{11}$ to transmitter 2 we also provide $W_{12}$ to transmitter 1. Now, both $W_{11}$ and $W_{12}$ are shared between the two transmitters and destined for the same receiver, so they can be combined into one message $W_0$ so that we obtain Fig. \ref{fig:xshared}(b). Since Scenario $2$ is obtained by providing $W_{12}$ to transmitter $1$ the total number of degrees of freedom for Scenario $2$ cannot be smaller than the $X$ channel with only one shared message $W_{11}$. Therefore,  the total number of degrees of freedom in the two scenarios are the same.

Note that the above argument also shows that, for the total degrees of freedom, sharing two messages (e.g. $W_{12}$ and $W_{11}$) is equivalent to sharing one message if both messages have the same destination.

The converse proof of Theorem \ref{theorem:cogxtransmitter} utilizes the result of Theorem \ref{theorem:cogint} while the achievability proof is similar to that for the MIMO $X$ channel (over a 2-symbol extension). Detailed proofs  are presented in the Appendix.


\subsection{MIMO $X$ channel with a Cognitive Receiver}
\begin{theorem}\label{theorem:cogxreceiver}
On the MIMO $X$ channel with $M>1$ antennas at all nodes if message $W_{11}$ is made available  to receiver $2$, the total number of degrees of freedom 
\begin{equation}
\eta^\star = \frac{3}{2}M.
\end{equation}
\end{theorem}
\begin{proof}(Converse)

We begin by finding outerbounds on the total degrees of freedom for the MIMO $X$ channel with a cognitive receiver shown in Fig. \ref{fig:cogxfull}(b). 
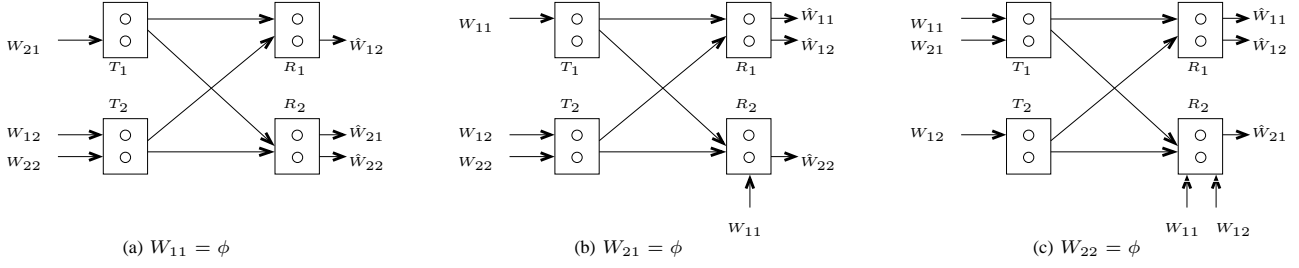
\begin{figure}[h]
\centerline{\input{cogxreceiver.pstex_t}}
\caption{Outerbounds for the Converse to Theorem \ref{theorem:cogxreceiver}.}\label{fig:cogxreceiver}
\end{figure}
\begin{enumerate}
\item Setting $W_{11}=\phi$, as shown in Fig. \ref{fig:cogxreceiver}(a), we obtain the outerbound $d_{12}+d_{21}+d_{22}\leq M$. The bound is straightforward because without $W_{11}$ there is no shared message and therefore the outerbounds on the regular MIMO $X$ channel apply.
\item Setting $W_{21}=\phi$, as shown in Fig. \ref{fig:cogxreceiver}(b), we obtain the outerbound $d_{11}+d_{12}+d_{22}\leq M$. Note that when $W_{21}=\phi$ and  $W_{11}$ is known to receiver $2$ then receiver $2$ is able to reconstruct and subtract transmitter $1$'s signal from its own received signal so that this case is equivalent to the $Z$ channel $Z(21)$ whose degrees of freedom were established in Corollary \ref{cor:z}. With equal number of antennas at all nodes, the $Z(21)$ channel has $M$ degrees of freedom, giving us our second outerbound.
\item Setting $W_{22}=\phi$ and also providing $W_{12}$ to receiver $2$, as shown in Fig. \ref{fig:cogxreceiver}(c), we obtain our third outerbound $d_{11}+d_{12}+d_{21}\leq M$ as follows. Start with any encoding scheme such that messages $W_{11}, W_{12}, W_{21}$ are reliably decoded at their intended receivers. Since receiver $2$ knows $W_{12}$ it can subtract the signal from transmitter $2$ from its received signal. Receiver $1$ by definition must be able to decode $W_{12}$ and therefore it can also subtract the signal from transmitter $2$ from its received signal. Now both receivers see only the signal from transmitter $1$ and both receivers know $W_{11}$ (Receiver $1$ knows $W_{11}$ because it must be able to decode it while receiver $2$ knows $W_{11}$ because of the cognitive assumption). Without affecting the degrees of freedom we can make the signal at receiver $1$ less noisy as in \cite{Jafar_dof_int} which would imply that if receiver $2$ can decode $W_{21}$ then so can receiver $1$. Thus receiver $1$ must be able to decode all three messages $W_{11}, W_{12}, W_{21}$ and the total degrees of freedom of these messages cannot be more than the number of antennas at receiver $1$. Thus we have our third outerbound.
\end{enumerate}
It may be an interesting exercise to verify that setting $W_{12}=\phi$ in the manner of the preceding three cases will not produce the outerbound $d_{11}+d_{21}+d_{22}\leq M$. 

Now let us try to maximize the total degrees of freedom subject to the three outerbounds obtained above.
\begin{eqnarray}
\eta^\star\leq d^\star&\define&\max{d_{11}+d_{12}+d_{21}+d_{22}}\\
\mbox{such that}&&d_{12}+d_{21}+d_{22}\leq M\\
&&d_{11}+d_{12}+d_{22}\leq M\\
&&d_{11}+d_{12}+d_{21}\leq M\\
&& d_{11}\geq0, d_{12}\geq 0, d_{21}\geq 0, d_{22}\geq 0.
\end{eqnarray}
The above is a linear program and we can write its dual problem as:
\begin{eqnarray}
d^\star&=&\min{(u_1+u_2+u_3)M}\\
\mbox{such that}&&u_1+u_2\geq 1\\
&& u_1+u_3\geq 1\\
&& u_1+u_2+u_3\geq 1\\
&& u_2+u_3\geq 1\\
&& u_1\geq 0, u_2\geq 0, u_3\geq 0.
\end{eqnarray}
Note that $u_1=u_2=u_3=0.5$ is a feasible point for the dual problem and therefore must be an upperbound on the primal problem. Substituting these values intothe dual objective we obtain
\begin{eqnarray*}
\eta^\star \leq \frac{3}{2}M.
\end{eqnarray*}
which is the converse result.
\end{proof}
Next we present the achievability proof.

\begin{proof}(Achievability)
Similar to the achievability proof of Theorem \ref{theorem:cogxtransmitter}, the achievability result of Theorem \ref{theorem:cogxreceiver} is based on a $2$ symbol extension of the channel. Since many arguments are similar and the proof of Theorem \ref{theorem:cogxtransmitter} is described in detail in the Appendix, we only present an outline of the proof here. The $2$ symbol extension  gives us an equivalent MIMO $X$ channel with $2M$ antennas at each node. We set $d_{11}=d_{21}=d_{22}=M$ and encode the message streams independently as usual. Using its $2M$ transmit antennas, transmitter $1$ sends $M$ dimensional coded signal for message $W_{21}$ along beamforming directions orthogonal to the first $M$ antennas of receiver $1$. Similarly transmitter $2$ sends the message $W_{22}$ along beamforming directions orthogonal to first $M$ antennas of receiver $1$ as well. Thus, with this interference alignment, the first $M$ antennas of receiver $1$ see no interference from any of the messages intended for receiver $2$, while these transmissions span a full $2M$ dimensional space at receiver $2$. Finally, transmitter $1$ sends message $W_{11}$ along $M$ dimensions that can be decoded without any interference by receiver $1$ using its first $M$ interference free antennas. Receiver $2$, on the other hand, is able to cancel all the interference due to message $W_{11}$ because of its cognitive knowledge of the message. In this manner we are able to achieve $3M$ degrees of freedom over a 2-symbol extension of the channel so that the achievability of $\frac{3}{2}M$ degrees of freedom is established.
\end{proof}

\section{Comparison of inner and outerbounds on the degrees of freedom region}\label{section:normalized}
The identical form of the innerbound $\mathcal{D}^X_{in}$ and the outerbound $\mathcal{D}^X_{out}$ on the degrees of freedom region for the MIMO $X$ channel is intriguing because the integer constraint for the innerbound is the only thing that sets them apart. In the previous sections we showed how achievability schemes can be extended to rational degrees of freedom by considering multi-letter extensions of the MIMO $X$ channel such that all rational values are scaled to integers. While the proof was provided for the case when all nodes have equal number of antennas, it is easy to verify that with the multi-letter extensions the innerbound and outerbound are tight in most cases. An important exception is the case $M=1$ where our achievable schemes do not benefit from multi-letter extensions. In this section, using $M=1$ as an example,  we evaluate the significance of the scenarios where the inner and outerbounds are not tight. First, we show that even in such cases, if we allow channel coefficients to vary over time/frequency slots the full degrees of freedom may be achieved. We illustrate this through a numerical example for the $M=1$ case. Second, we argue that the inner and outerbounds are in fact identical subject to a natural normalization. 

\subsection{Achievability of the full $4/3$ degrees of freedom for the $M=1$ case}\label{sec:timevarying}
From the preceding sections we know for this channel $1\leq\eta_X^\star\leq \frac{4}{3}$, i.e. the upper and lower bounds are not tight if the channel coefficients are fixed throughout the duration of communication. In this section, we show that if the channel coefficients vary over time/frequency, then the full $4/3$ degrees of freedom are achieved over the $3$ symbol extension of the channel model.
The input output equation for each orthogonal dimension can be collectively written in the vector form as a MIMO $X$ channel with $3$ antennas at all nodes and diagonal channels.
\begin{eqnarray*}
\underbrace{\left[\begin{array}{c}y^{[i]}(1)\\y^{[i]}(2)\\y^{[i]}(3)\end{array}\right]}_{{\bf Y}^{[i]}}=\sum_{j=1}^2\underbrace{\left[\begin{array}{ccc} h^{[ij]}(1)&0&0\\0&h^{[ij]}(2)&0\\0&0&h^{[ij]}(3)\end{array}\right]}_{{\bf H}^{[ij]}}\underbrace{\left[\begin{array}{c}x^{[j]}(1)\\x^{[j]}(2)\\x^{[j]}(3)\end{array}\right]}_{{\bf X}^{[j]}}+\underbrace{\left[\begin{array}{c}n^{[i]}(1)\\n^{[i]}(2)\\n^{[i]}(3)\end{array}\right]}_{{\bf N}^{[i]}}, ~~i\in\{1,2\}.
\end{eqnarray*}
where, over the $k^{th}$ orthogonal frequency (time) slot, $y(i)(k), h^{[ij]}(k), x^{[j]}(k), n^{[i]}(k)$ represent the output of receiver $i$, the channel between transmitter $j$ and receiver $i$, the input signal from transmitter $j$ and the noise at receiver $i$. However, note that if the channel coefficients are drawn from a continuous joint distribution then almost surely the extended channel matrices satisfy the conditions of Theorem \ref{theorem:noninteger}. In other words the matrices are nonsingular and ${\bf F}$ has distinct eigenvalues. In fact, Theorem \ref{theorem:noninteger} implies that if any two of the three eigenvalues are distinct then this 3-symbol extended channel will achieve $4$ degrees of freedom. It is interesting to note that we do not need all channel coefficients to vary. It suffices to have any one of the channel coefficients takes at least one unique value over three time/frequency slots, i.e., even if one channel coefficient is time varying the full $4/3$ degrees of freedom are almost surely achieved on the MIMO $X$ channel with one antenna at each node.


Next we show that the inner and outerbounds on the degrees of freedom are always tight subject to a natural normalization.
\subsection{Normalized Degrees of Freedom}
Consider the point to point MIMO channel with $M$ transmit antennas and $N$ receive antennas. The degrees of freedom for this channel are well known to be $\min(N,M)$. Now, suppose we scale the number of transmit and receive antennas by the same positive integer constant $\kappa\in\mathbb{Z}_+$ so that we have $\kappa M$ transmit antennas and $\kappa N$ receive antennas. Then, the degrees of freedom also scale by $\kappa$ as $\min(\kappa N, \kappa M)=\kappa \min(N,M)$. The same property can be observed for the multiple access channel, where $\min(\kappa N, \kappa M_1+\kappa M_2)=\kappa\min(N,M_1+M_2)$, the broadcast channel, where $\min(\kappa N_1+\kappa N_2, \kappa M)=\kappa\min(N_1+N_2,M)$ and the interference channel where $\min(\kappa M_1+\kappa M_2, \kappa N_1+\kappa N_2,\max(\kappa M_1, \kappa N_2),\max(\kappa M_2, \kappa N_1))=\kappa \min( M_1+ M_2,  N_1+ N_2,\max(M_1,  N_2),\max(M_2,N_1))$. So, for all these channels, scaling the number of antennas at every node by the same factor amounts to scaling the degrees of freedom by the same factor as well. 

This prompts the normalized definition of degrees of freedom region for the MIMO $X$ channel as follows:
\begin{eqnarray*}
\overline{\mathcal{D}}^{X(N_1,N_2,M_1,M_2)}\define\{(d_{11},d_{12},d_{21},d_{22}):\exists \kappa\in\mathbb{Z}_+ ~~\mbox{for which} ~~(\kappa d_{11}, \kappa d_{12}, \kappa d_{21}, \kappa d_{22})\in\mathcal{D}^{X(\kappa N_1, \kappa N_2, \kappa M_1,\kappa M_2)}\}
\end{eqnarray*}
where $X(N_1,N_2,M_1,M_2)$ refers to the MIMO $X$ channel with $N_1,N_2$ receive antennas and $M_1,M_2$ transmit antennas.

The normalized definition is especially important when other dimensions such as time and frequency are involved in addition to space. Consider the $(M,N)$ point to point channel again. If there are $\kappa$ orthogonal dimensions available in the form of orthogonal frequency bands or blocks of time, then these orthogonal dimensions can be represented in space in the form of a total of $\kappa M$ transmit antennas and a total of $\kappa N$ receive antennas giving rise to a block diagonal channel matrix of size $\kappa N\times \kappa M$ which would yield a total of $\kappa\min(N,M)$ degrees of freedom. The normalized definition of degrees of freedom is especially relevant in this case as it captures the \emph{rate} at which the degrees of freedom scale with orthogonal dimensions.

In a similar manner, we can normalize the inner and outerbounds so that:
\begin{eqnarray*}
\overline{\mathcal{D}}^{X(N_1,N_2,M_1,M_2)}_{out}&\define&\{(d_{11},d_{12},d_{21},d_{22}):\exists \kappa\in\mathbb{Z}_+ ~~\mbox{for which} ~~(\kappa d_{11}, \kappa d_{12}, \kappa d_{21}, \kappa d_{22})\in\mathcal{D}^{X(\kappa N_1, \kappa N_2, \kappa M_1,\kappa M_2)}_{out} \}\\
\overline{\mathcal{D}}^{X(N_1,N_2,M_1,M_2)}_{in}&\define&\{(d_{11},d_{12},d_{21},d_{22}):\exists \kappa\in\mathbb{Z}_+ ~~\mbox{for which} ~~(\kappa d_{11}, \kappa d_{12}, \kappa d_{21}, \kappa d_{22})\in\mathcal{D}^{X(\kappa N_1, \kappa N_2, \kappa M_1,\kappa M_2)}_{in}\}
\end{eqnarray*}

\begin{lemma}\label{lemma:outequalsout}
$\overline{\mathcal{D}}^X_{out}={\mathcal{D}}^X_{out}$.
\end{lemma}
\begin{proof}
This is easily verified because all the inequalities that define ${\mathcal{D}}^X_{out}$ are unaffected by scaling with $\kappa$.
\end{proof}
\begin{lemma}\label{lemma:inequalsout}
$\overline{\mathcal{D}}^X_{in}={\mathcal{D}}^X_{out}$.
\end{lemma}
\begin{proof}
Because ${\mathcal{D}}^X_{out}$ is bounded and defined by linear inequalities, it is a convex polyhedron. $\overline{\mathcal{D}}^X_{in}$ is also a convex set because of the convex hull operation. In order to prove that $\overline{\mathcal{D}}^X_{in}={\mathcal{D}}^X_{out}$ it suffices to show that all the extreme points of the convex polyhedron ${\mathcal{D}}^X_{out}$ are contained within $\overline{\mathcal{D}}^X_{in}$. We show this as follows.

${\mathcal{D}}^X_{out}$ is a convex polyhedron in $4$ dimensions, defined by the linear inequalities which form its bounding hyperplanes (including the positivity constraints $d_{ij}\geq 0$). By the fundamental theorem of linear programming, all of the extreme points of this convex polyhedron are uniquely defined by the intersection of at least $4$ of these hyperplanes, i.e. where at least $4$ of these inequalities are satisfied with an equality. Now, because the linear inequalities have all rational (in fact, integer) coefficients and constants, all the extreme points of ${\mathcal{D}}^X_{out}$ have rational coefficients as well. Without loss of generality we can represent any extreme point of ${\mathcal{D}}^X_{out}$ as:
\begin{eqnarray*}
(d_{11},d_{12}, d_{21},d_{22})=\left(\frac{p_{11}}{q_{11}},  \frac{p_{12}}{q_{12}}, \frac{p_{21}}{q_{21}}, \frac{p_{22}}{q_{22}}\right)
\end{eqnarray*}
where $p_{ij}, q_{ij}$ are all integers and have no common factors.
Now, let us choose $\kappa=LCM(q_{11},q_{12},q_{21},q_{22})$ where $LCM$ stands for the least common multiple. Then clearly $(\kappa d_{11}, \kappa d_{12}, \kappa d_{21}, \kappa d_{22})\in\mathcal{D}^{X(\kappa N_1, \kappa N_2, \kappa M_1,\kappa M_2)}_{in}$ because $(\kappa d_{11}, \kappa d_{12}, \kappa d_{21}, \kappa d_{22})$ are integers that satisfy the linear inqualities in $\mathcal{D}^{X(\kappa N_1, \kappa N_2, \kappa M_1,\kappa M_2)}_{out}$ which are the same inequalities as in $\mathcal{D}^{X(\kappa N_1, \kappa N_2, \kappa M_1,\kappa M_2)}_{in}$. Thus, all extreme points of ${\mathcal{D}}^X_{out}$ are contained in ${\mathcal{D}}^X_{in}$. Since the former is an outerbound, the two must be equal and we have the result of Lemma \ref{lemma:inequalsout}.
\end{proof}
The preceding lemmas lead us to the main result of this section, which is the precise characterization of the normalized degrees of freedom for the MIMO $X$ channel.
\begin{theorem}\label{theorem:normal}
$\overline{\mathcal{D}}^X={\mathcal{D}}^X_{out}$, and $\overline{\eta}^\star_X = {\eta}_{out}.$
\end{theorem}
\begin{proof}
\begin{eqnarray*}
\overline{\mathcal{D}}^X&\subset&\overline{\mathcal{D}}^X_{out}={\mathcal{D}}^X_{out}~~~(\mbox{Lemma } \ref{lemma:outequalsout})\\
\overline{\mathcal{D}}^X&\supset&\overline{\mathcal{D}}^X_{in}={\mathcal{D}}^X_{out}~~~(\mbox{Lemma }\ref{lemma:inequalsout})\\
\Rightarrow \overline{\mathcal{D}}^X&=&{\mathcal{D}}^X_{out}.\\
\Rightarrow \overline{\eta}^\star_X=\max_{\overline{\mathcal{D}}^X}(d_{11}+d_{12}+d_{21}+d_{22})&=&\max_{{\mathcal{D}}^X_{out}}(d_{11}+d_{12}+d_{21}+d_{22}){\mathcal{D}}^X_{out}=\eta_{out}.
\end{eqnarray*}
\end{proof}

\section{Conclusion}
We characterize the degrees of freedom region for the MIMO $X$ channel, a system with two multiple antenna transmitters and two multiple antenna receivers where independent messages are communicated from each transmitter to each receiver. The $X$ channel is especially interesting because the interference channel, the multiple access channel and the broadcast channels are special cases of this channel. Studying the $X$ channel brings out some interesting aspects of multiuser MIMO communications that are not revealed by studying the individual MAC, BC and interference channel components. 

First is the idea of interference alignment. A powerful idea on the MIMO $X$ channel, interference alignment is not needed on the BC, MAC and interference channels with full channel knowledge\footnote{When channel knowldege is not available, as in the compound broadcast channel, interference alignment may be quite useful \cite{Weingarten_Shamai_Kramer}.}. While zero forcing has been the universal scheme to achieve the full degrees of freedom for the MIMO MAC, BC and interference channels, the $X$ channel provides us another fundamental tool in the form of interference alignment that can be combined with zero forcing to estimate the degrees of freedom of wireless networks.

The second interesting aspect of the $X$ channel is that unlike the MIMO MAC, BC and interference channels, it can have fractional degrees of freedom. In fact, to the best of our knowledge, the $X$ channel is the first multiuser communications scenario with non-degenerate channels and full channel knowledge where fractional degrees of freedom have been shown to be optimal. The way the fractional degrees of freedom are achieved is also insightful. We are able to construct achievability schemes over the $K$-letter extension of the channel. The value of $K$ is chosen to be the scaling factor that converts all rational degrees of freedom terms to integer values. 

The significance of the $K$-letter extensions in our achievability schemes provides another reason to study the degrees of freedom for MIMO nodes with large number of antennas at each node. A channel model with a large number of antennas at each node may seem to be of limited practical significance. However, note that even if a network consists of only one antenna at each node, its $K$-symbol extension converts each node into a $K$ antenna node with a diagonalized channel matrix structure. As we saw in this work, the degrees of freedom characterizations for MIMO nodes with $KM$ antennas at each node also apply directly to the $K$ symbol extensions of the same network with $M$ antennas at each node. Studying MIMO networks with large $KM$ allows us to find optimal spatial alignments that can directly translate into temporal alignments of precoding vectors over multiletter extensions of the channel. While the multiletter channel extensions may not be essential from a capacity perspective, evidently they can significantly simplify the achievable scheme to a linear preprocessing scheme at the transmitters.

Finally, the $X$ channel is also interesting in how it is able to benefit from cognitive message sharing. With equal number ($M>1$) of antennas at all nodes, we found that the number of degrees of freedom on the MIMO interference channel does not benefit from a shared message, regardless of whether the message is shared with the other user's transmitter, receiver or both. However, on the MIMO $X$ channel even one shared message increases the number of degrees of freedom from $\frac{4}{3}M$ to $\frac{3}{2}M$. The increase in the degrees of freedom from cognitive cooperation is the same regardless of whether the message sharing occurs through a cognitive transmitter or a cognitive receiver. This duality   between cognitive transmitter and cognitive receiver message sharing (from a degrees of freedom perspective) is an interesting feature of both the $X$ channel and the interference channel. 




\section{Appendix}
\subsection{Proof of Theorem \ref{theorem:cogxtransmitter}: Achievability}\label{app:ach}
For this proof we assume that the channel matrices ${\bf H}^{[11]}, {\bf H}^{[12]},{\bf H}^{[21]},{\bf H}^{[22]}$  are invertible and the product channel matrix:
\begin{equation}
{\bf F}\define\left( {\bf H}^{[11]}\right)^{-1} {\bf H}^{[12]}\left( {\bf H}^{[22]}\right)^{-1}{\bf H}^{[21]}
\end{equation}
has $M$ distinct eigenvalues. Both conditions are satisfied with probability $1$ when the channel matrices are generated according to a continuous probability distribution.

The achievability proof is similar to Theorem \ref{theorem:noninteger} with some interesting differences that we highlight here. Instead of a 3-symbol extension channel as in Theorem \ref{theorem:noninteger}, here we consider a 2-symbol extension channel so that all the channel matrices $\overline{\bf H}^{[ij]}$ are $2M\times 2M$ block diagonal matrices formed by repeating the original channel matrices ${\bf H}^{[ij]}$ twice along the main diagonal. We assign equal number of degrees of freedom $d_{11}=d_{21}=d_{22}=M$ to the three messages $W_{11}, W_{21}, W_{22}$ and zero degrees of freedom ($d_{12}=0$) to $W_{12}$. Instead of transmitting the encoded streams ${\bf X}^{[12]}$ for message $W_{12}$ along $\overline{\bf V}^{[12]}$ we transmit the encoded streams for message $W_{11}$ which is available to the second transmitter because of message sharing. This is accomplished by setting
\begin{eqnarray}
{\bf X}^{[12]}&=&-{\bf X}^{[11]}. \label{eq:signalalign}
\end{eqnarray}
The transmit directions are picked exactly as in (\ref{eq:transmitdirection1}) and (\ref{eq:transmitdirection2}), so that the $2M\times 1$  received signals at the two receivers are
\begin{eqnarray}
\overline{\bf Y}^{[1]}&=&\left(\overline{\bf H}^{[11]}-\overline{\bf H}^{[12]}\left(\overline{\bf H}^{[22]}\right)^{-1}\overline{\bf H}^{[21]}\right)\overline{\bf V}^{[11]}{\bf X}^{[11]}+\overline{\bf H}^{[11]}\overline{\bf V}^{[21]}\left({\bf X}^{[21]}+{\bf X}^{[22]}\right)+\overline{\bf N}^{[1]}\\
\overline{\bf Y}^{[2]}&=&\overline{\bf H}^{[21]}\overline{\bf V}^{[21]}{\bf X}^{[21]}+\overline{\bf H}^{[22]} \left(\overline{\bf H}^{[12]}\right)^{-1}\overline{\bf H}^{[11]}\overline{\bf V}^{[21]}{\bf X}^{[22]}+\overline{\bf N}^{[2]}
\end{eqnarray}
As before, the interference from the signals ${\bf X}^{[21]}$ and ${\bf X}^{[22]}$ overlaps at receiver 1. However, interestingly all the interference is eliminated at receiver 2 by construction according to (\ref{eq:signalalign}), (\ref{eq:transmitdirection1}) and (\ref{eq:transmitdirection2}). Intuitively, this is possible because both transmitter 1 and transmitter 2 know message $W_{11}$ and together they have $2M$ transmit antennas which allows them to zero force the common signal at all $M$ antennas of receiver $2$. Now, as before, receiver 1 can separate desired signal ${\bf X}^{[11]}$ from interference signals ${\bf X}^{[21]}+{\bf X}^{[22]}$ if all signal and interference directions are independent. In other words we require
\begin{eqnarray}
\mbox{rank}\left[\left(\overline{\bf H}^{[11]}-\overline{\bf H}^{[12]}\left(\overline{\bf H}^{[22]}\right)^{-1}\overline{\bf H}^{[21]}\right)\overline{\bf V}^{[11]}~~~~\overline{\bf H}^{[11]}\overline{\bf V}^{[21]}\right]=\mbox{rank}\left[\left({\bf I}_{2M\times 2M}-\overline{\bf F}\right)\overline{\bf V}^{[11]} ~~~~\overline{\bf V}^{[21]}\right]=2M\label{eq:rankcondition1}\\
\mbox{rank}\left[\overline{\bf H}^{[21]}\overline{\bf V}^{[21]}~~~~\overline{\bf H}^{[22]} \left(\overline{\bf H}^{[12]}\right)^{-1}\overline{\bf H}^{[11]}\overline{\bf V}^{[21]}\right]=\mbox{rank}\left[\overline{\bf V}^{[21]}~~~  \overline{\bf F}^{-1}\overline{\bf V}^{[21]} \right] =2M\label{eq:rankcondition2}
\end{eqnarray}
The first condition (\ref{eq:rankcondition1}) is easily satisfied because for any choice of ${\bf V}^{[21]}$ that satisfies (\ref{eq:rankcondition2}) we can easily choose ${\bf V}^{[11]}$ and it does not affect the second condition (\ref{eq:rankcondition2}). To find a ${\bf V}^{[21]}$ that satisfies (\ref{eq:rankcondition2}) we first align the coordinate system along the eigenbasis ${\bf E}_F$ of ${\bf F}$ as defined in (\ref{eq:eigenbasis}) so that $\overline{\bf V}^{[21]}=\overline{\bf E}_F{\bf V}^{[21]}_F$. Instead of grouping the eigenvalues as before, we simply assume (without loss of generality) that $\overline\Lambda$ is a diagonal matrix with the vector $[\lambda_1,\cdots,\lambda_M,\lambda_1,\cdots,\lambda_M]$ along the main diagonal. We then choose 
\begin{eqnarray*}
{\bf V}^{[21]}_F=\left[\begin{array}{c}{\bf I}_{M\times M}\\{\overleftarrow{\bf I}}_{M\times M}\end{array}\right].
\end{eqnarray*}
With this choice of ${\bf V}^{[21]}$ we show that 
\begin{eqnarray*}
\left[{\bf V}^{[21]}_F  ~~~\overline\Lambda^{-1}{\bf V}^{[21]}_F\right]{\bf A}={\bf 0} &\mbox{only if}& {\bf A}={\bf 0}
\end{eqnarray*}
Rows $i$ and $M+1+(i)\mbox{mod}(M)$ of $\left[{\bf V}^{[21]}_F  ~~~\overline\Lambda^{-1}{\bf V}^{[21]}_F\right]{\bf A}={\bf 0}$ provide the equations:
\begin{eqnarray*}
a_i+\frac{1}{\lambda_i}a_{M+i}&=&0\\
a_i+\frac{1}{\lambda_{1+(i)\mbox{mod}(M)}}a_{M+i}&=&0.
\end{eqnarray*}
Since $\lambda_i\neq\lambda_{1+(i)\mbox{mod}(M)}$, we must have $a_i=a_{M+i}=0$ for all $i\in\{1,2,\cdots,M\}$ which means rank$\left[{\bf V}^{[21]}_F  ~~~\overline\Lambda^{-1}{\bf V}^{[21]}_F \right]=2M$. Thus $M$ degrees of freedom are achieved by each of the messages $W_{11},W_{21},W_{22}$. Since $3M$ degrees of freedom are achieved over the 2-symbol extension channel model, we have established the achievability of $\frac{3}{2}M$ degrees of freedom.

\subsection{Proof of Theorem \ref{theorem:cogxtransmitter}: Converse} \label{app:conv}

Using Theorem \ref{theorem:cogint} we prove the converse for the equivalent channel model in Fig. \ref{fig:xshared}(b) as follows:
\begin{eqnarray}
\max\lim_{\rho\rightarrow\infty} \frac{R_0(\rho)+R_1(\rho)}{\log(\rho)} &\leq & M\\
\max\lim_{\rho\rightarrow\infty} \frac{R_0(\rho)+R_2(\rho)}{\log(\rho)} &\leq & M\\
\max\lim_{\rho\rightarrow\infty} \frac{R_1(\rho)+R_2(\rho)}{\log(\rho)} &\leq & M
\end{eqnarray}
\begin{eqnarray}
\eta=\max\lim_{\rho\rightarrow\infty} \frac{R_0(\rho)+R_1(\rho)+R_2(\rho)}{\log(\rho)} &\leq& \frac{3}{2}M.\nonumber
\end{eqnarray}
In the above derivation, the first two bounds follow from Theorem \ref{theorem:cogint}, the third bound follows from the interference channel \cite{Jafar_dof_int} and the last bound is obtained by summing the first three bounds to give us the result.

\bibliographystyle{ieeetr}
\bibliography{Thesis}
\end{document}

%% file: xchannel.pstex_t
\begin{picture}(0,0)%
\includegraphics{xchannel.pstex}%
\end{picture}%
\setlength{\unitlength}{2408sp}%
\begingroup\makeatletter\ifx\SetFigFont\undefined%
\gdef\SetFigFont#1#2#3#4#5{%
  \reset@font\fontsize{#1}{#2pt}%
  \fontfamily{#3}\fontseries{#4}\fontshape{#5}%
  \selectfont}%
\fi\endgroup%
\begin{picture}(5466,2766)(268,-2194)
\put(4126,314){\makebox(0,0)[lb]{\smash{\SetFigFont{7}{8.4}{\rmdefault}{\mddefault}{\updefault}\special{ps: gsave 0 0 0 setrgbcolor}$R_1$\special{ps: grestore}}}}
\put(1801,314){\makebox(0,0)[lb]{\smash{\SetFigFont{7}{8.4}{\rmdefault}{\mddefault}{\updefault}\special{ps: gsave 0 0 0 setrgbcolor}$T_1$\special{ps: grestore}}}}
\put(1801,-1261){\makebox(0,0)[lb]{\smash{\SetFigFont{7}{8.4}{\rmdefault}{\mddefault}{\updefault}\special{ps: gsave 0 0 0 setrgbcolor}$T_2$\special{ps: grestore}}}}
\put(4126,-1261){\makebox(0,0)[lb]{\smash{\SetFigFont{7}{8.4}{\rmdefault}{\mddefault}{\updefault}\special{ps: gsave 0 0 0 setrgbcolor}$R_2$\special{ps: grestore}}}}
\put(451, 14){\makebox(0,0)[lb]{\smash{\SetFigFont{7}{8.4}{\rmdefault}{\mddefault}{\updefault}\special{ps: gsave 0 0 0 setrgbcolor}$W_{11}$\special{ps: grestore}}}}
\put(451,-361){\makebox(0,0)[lb]{\smash{\SetFigFont{7}{8.4}{\rmdefault}{\mddefault}{\updefault}\special{ps: gsave 0 0 0 setrgbcolor}$W_{21}$\special{ps: grestore}}}}
\put(451,-1561){\makebox(0,0)[lb]{\smash{\SetFigFont{7}{8.4}{\rmdefault}{\mddefault}{\updefault}\special{ps: gsave 0 0 0 setrgbcolor}$W_{12}$\special{ps: grestore}}}}
\put(451,-1861){\makebox(0,0)[lb]{\smash{\SetFigFont{7}{8.4}{\rmdefault}{\mddefault}{\updefault}\special{ps: gsave 0 0 0 setrgbcolor}$W_{22}$\special{ps: grestore}}}}
\put(5101,-361){\makebox(0,0)[lb]{\smash{\SetFigFont{7}{8.4}{\rmdefault}{\mddefault}{\updefault}\special{ps: gsave 0 0 0 setrgbcolor}$\hat{W}_{12}$\special{ps: grestore}}}}
\put(5101,-61){\makebox(0,0)[lb]{\smash{\SetFigFont{7}{8.4}{\rmdefault}{\mddefault}{\updefault}\special{ps: gsave 0 0 0 setrgbcolor}$\hat{W}_{11}$\special{ps: grestore}}}}
\put(5101,-1561){\makebox(0,0)[lb]{\smash{\SetFigFont{7}{8.4}{\rmdefault}{\mddefault}{\updefault}\special{ps: gsave 0 0 0 setrgbcolor}$\hat{W}_{21}$\special{ps: grestore}}}}
\put(5101,-1936){\makebox(0,0)[lb]{\smash{\SetFigFont{7}{8.4}{\rmdefault}{\mddefault}{\updefault}\special{ps: gsave 0 0 0 setrgbcolor}$\hat{W}_{22}$\special{ps: grestore}}}}
\end{picture}

%% file: xalign.pstex_t
\begin{picture}(0,0)%
\includegraphics{xalign.pstex}%
\end{picture}%
\setlength{\unitlength}{2408sp}%
\begingroup\makeatletter\ifx\SetFigFont\undefined%
\gdef\SetFigFont#1#2#3#4#5{%
  \reset@font\fontsize{#1}{#2pt}%
  \fontfamily{#3}\fontseries{#4}\fontshape{#5}%
  \selectfont}%
\fi\endgroup%
\begin{picture}(8391,4341)(2368,-3469)
\put(5926,-1111){\makebox(0,0)[lb]{\smash{\SetFigFont{7}{8.4}{\rmdefault}{\mddefault}{\updefault}\special{ps: gsave 0 0 0 setrgbcolor}${\bf H}^{[12]}$\special{ps: grestore}}}}
\put(5926,-1936){\makebox(0,0)[lb]{\smash{\SetFigFont{7}{8.4}{\rmdefault}{\mddefault}{\updefault}\special{ps: gsave 0 0 0 setrgbcolor}${\bf H}^{[21]}$\special{ps: grestore}}}}
\put(5401,-2611){\makebox(0,0)[lb]{\smash{\SetFigFont{7}{8.4}{\rmdefault}{\mddefault}{\updefault}\special{ps: gsave 0 0 0 setrgbcolor}${\bf H}^{[22]}$\special{ps: grestore}}}}
\put(5401,-511){\makebox(0,0)[lb]{\smash{\SetFigFont{7}{8.4}{\rmdefault}{\mddefault}{\updefault}\special{ps: gsave 0 0 0 setrgbcolor}${\bf H}^{[11]}$\special{ps: grestore}}}}
\put(9001,-361){\makebox(0,0)[lb]{\smash{\SetFigFont{7}{8.4}{\rmdefault}{\mddefault}{\updefault}\special{ps: gsave 0 0 0 setrgbcolor}${\bf H}^{[11]}{\bf U}^{[11]}{\bf X}^{[11]}$\special{ps: grestore}}}}
\put(7351,-1036){\makebox(0,0)[lb]{\smash{\SetFigFont{7}{8.4}{\rmdefault}{\mddefault}{\updefault}\special{ps: gsave 0 0 0 setrgbcolor}${\bf H}^{[12]}{\bf U}^{[12]}{\bf X}^{[12]}$\special{ps: grestore}}}}
\put(7351,-3286){\makebox(0,0)[lb]{\smash{\SetFigFont{7}{8.4}{\rmdefault}{\mddefault}{\updefault}\special{ps: gsave 0 0 0 setrgbcolor}${\bf H}^{[21]}{\bf U}^{[21]}{\bf X}^{[21]}$\special{ps: grestore}}}}
\put(9076,-2611){\makebox(0,0)[lb]{\smash{\SetFigFont{7}{8.4}{\rmdefault}{\mddefault}{\updefault}\special{ps: gsave 0 0 0 setrgbcolor}${\bf H}^{[22]}{\bf U}^{[22]}{\bf X}^{[22]}$\special{ps: grestore}}}}
\put(8176,614){\makebox(0,0)[lb]{\smash{\SetFigFont{7}{8.4}{\rmdefault}{\mddefault}{\updefault}\special{ps: gsave 0 0 0 setrgbcolor}${\bf H}^{[12]}{\bf U}^{[22]}{\bf X}^{[22]}$\special{ps: grestore}}}}
\put(8176,239){\makebox(0,0)[lb]{\smash{\SetFigFont{7}{8.4}{\rmdefault}{\mddefault}{\updefault}\special{ps: gsave 0 0 0 setrgbcolor}${\bf H}^{[11]}{\bf U}^{[21]}{\bf X}^{[21]}$\special{ps: grestore}}}}
\put(8251,-2011){\makebox(0,0)[lb]{\smash{\SetFigFont{7}{8.4}{\rmdefault}{\mddefault}{\updefault}\special{ps: gsave 0 0 0 setrgbcolor}${\bf H}^{[21]}{\bf U}^{[11]}{\bf X}^{[11]}$\special{ps: grestore}}}}
\put(8251,-1636){\makebox(0,0)[lb]{\smash{\SetFigFont{7}{8.4}{\rmdefault}{\mddefault}{\updefault}\special{ps: gsave 0 0 0 setrgbcolor}${\bf H}^{[22]}{\bf U}^{[12]}{\bf X}^{[12]}$\special{ps: grestore}}}}
\put(2551, 89){\makebox(0,0)[lb]{\smash{\SetFigFont{7}{8.4}{\rmdefault}{\mddefault}{\updefault}\special{ps: gsave 0 0 0 setrgbcolor}${\bf U}^{[11]}{\bf X}^{[11]}$\special{ps: grestore}}}}
\put(2476,-2161){\makebox(0,0)[lb]{\smash{\SetFigFont{7}{8.4}{\rmdefault}{\mddefault}{\updefault}\special{ps: gsave 0 0 0 setrgbcolor}${\bf U}^{[12]}{\bf X}^{[12]}$\special{ps: grestore}}}}
\put(3226,-1936){\makebox(0,0)[lb]{\smash{\SetFigFont{7}{8.4}{\rmdefault}{\mddefault}{\updefault}\special{ps: gsave 0 0 0 setrgbcolor}${\bf U}^{[22]}{\bf X}^{[22]}$\special{ps: grestore}}}}
\put(3301,-136){\makebox(0,0)[lb]{\smash{\SetFigFont{7}{8.4}{\rmdefault}{\mddefault}{\updefault}\special{ps: gsave 0 0 0 setrgbcolor}${\bf U}^{[21]}{\bf X}^{[21]}$\special{ps: grestore}}}}
\end{picture}

%% file: intcogtxrx1.pstex_t
\begin{picture}(0,0)%
\includegraphics{intcogtxrx1.pstex}%
\end{picture}%
\setlength{\unitlength}{2408sp}%
\begingroup\makeatletter\ifx\SetFigFont\undefined%
\gdef\SetFigFont#1#2#3#4#5{%
  \reset@font\fontsize{#1}{#2pt}%
  \fontfamily{#3}\fontseries{#4}\fontshape{#5}%
  \selectfont}%
\fi\endgroup%
\begin{picture}(13242,2787)(421,-2386)
\put(4086,-164){\makebox(0,0)[lb]{\smash{\SetFigFont{5}{6.0}{\rmdefault}{\mddefault}{\updefault}\special{ps: gsave 0 0 0 setrgbcolor}$\hat{W}_{11}$\special{ps: grestore}}}}
\put(4086,-1362){\makebox(0,0)[lb]{\smash{\SetFigFont{5}{6.0}{\rmdefault}{\mddefault}{\updefault}\special{ps: gsave 0 0 0 setrgbcolor}$ \hat{W}_{22}$\special{ps: grestore}}}}
\put(5035,-164){\makebox(0,0)[lb]{\smash{\SetFigFont{5}{6.0}{\rmdefault}{\mddefault}{\updefault}\special{ps: gsave 0 0 0 setrgbcolor}$W_{11}$\special{ps: grestore}}}}
\put(5035,-1362){\makebox(0,0)[lb]{\smash{\SetFigFont{5}{6.0}{\rmdefault}{\mddefault}{\updefault}\special{ps: gsave 0 0 0 setrgbcolor}$W_{22}$\special{ps: grestore}}}}
\put(8574,-164){\makebox(0,0)[lb]{\smash{\SetFigFont{5}{6.0}{\rmdefault}{\mddefault}{\updefault}\special{ps: gsave 0 0 0 setrgbcolor}$\hat{W}_{11}$\special{ps: grestore}}}}
\put(8574,-1362){\makebox(0,0)[lb]{\smash{\SetFigFont{5}{6.0}{\rmdefault}{\mddefault}{\updefault}\special{ps: gsave 0 0 0 setrgbcolor}$ \hat{W}_{22}$\special{ps: grestore}}}}
\put(547,-1362){\makebox(0,0)[lb]{\smash{\SetFigFont{5}{6.0}{\rmdefault}{\mddefault}{\updefault}\special{ps: gsave 0 0 0 setrgbcolor}$W_{22}$\special{ps: grestore}}}}
\put(2176,-2386){\makebox(0,0)[lb]{\smash{\SetFigFont{7}{8.4}{\rmdefault}{\mddefault}{\updefault}\special{ps: gsave 0 0 0 setrgbcolor}(a)\special{ps: grestore}}}}
\put(11326,-2386){\makebox(0,0)[lb]{\smash{\SetFigFont{7}{8.4}{\rmdefault}{\mddefault}{\updefault}\special{ps: gsave 0 0 0 setrgbcolor}(c)\special{ps: grestore}}}}
\put(6826,-2386){\makebox(0,0)[lb]{\smash{\SetFigFont{7}{8.4}{\rmdefault}{\mddefault}{\updefault}\special{ps: gsave 0 0 0 setrgbcolor}(b)\special{ps: grestore}}}}
\put(12076,-2086){\makebox(0,0)[lb]{\smash{\SetFigFont{5}{6.0}{\rmdefault}{\mddefault}{\updefault}\special{ps: gsave 0 0 0 setrgbcolor}$W_{11}$\special{ps: grestore}}}}
\put(7801,-2086){\makebox(0,0)[lb]{\smash{\SetFigFont{5}{6.0}{\rmdefault}{\mddefault}{\updefault}\special{ps: gsave 0 0 0 setrgbcolor}$W_{11}$\special{ps: grestore}}}}
\put(12151,-511){\makebox(0,0)[lb]{\smash{\SetFigFont{5}{6.0}{\rmdefault}{\mddefault}{\updefault}\special{ps: gsave 0 0 0 setrgbcolor}$R_1$\special{ps: grestore}}}}
\put(10426,-511){\makebox(0,0)[lb]{\smash{\SetFigFont{5}{6.0}{\rmdefault}{\mddefault}{\updefault}\special{ps: gsave 0 0 0 setrgbcolor}$T_1$\special{ps: grestore}}}}
\put(7876,-511){\makebox(0,0)[lb]{\smash{\SetFigFont{5}{6.0}{\rmdefault}{\mddefault}{\updefault}\special{ps: gsave 0 0 0 setrgbcolor}$R_1$\special{ps: grestore}}}}
\put(6076,-511){\makebox(0,0)[lb]{\smash{\SetFigFont{5}{6.0}{\rmdefault}{\mddefault}{\updefault}\special{ps: gsave 0 0 0 setrgbcolor}$T_1$\special{ps: grestore}}}}
\put(1576,-511){\makebox(0,0)[lb]{\smash{\SetFigFont{5}{6.0}{\rmdefault}{\mddefault}{\updefault}\special{ps: gsave 0 0 0 setrgbcolor}$T_1$\special{ps: grestore}}}}
\put(3376,-511){\makebox(0,0)[lb]{\smash{\SetFigFont{5}{6.0}{\rmdefault}{\mddefault}{\updefault}\special{ps: gsave 0 0 0 setrgbcolor}$R_1$\special{ps: grestore}}}}
\put(3376,-886){\makebox(0,0)[lb]{\smash{\SetFigFont{5}{6.0}{\rmdefault}{\mddefault}{\updefault}\special{ps: gsave 0 0 0 setrgbcolor}$R_2$\special{ps: grestore}}}}
\put(1576,-886){\makebox(0,0)[lb]{\smash{\SetFigFont{5}{6.0}{\rmdefault}{\mddefault}{\updefault}\special{ps: gsave 0 0 0 setrgbcolor}$T_2$\special{ps: grestore}}}}
\put(12151,-886){\makebox(0,0)[lb]{\smash{\SetFigFont{5}{6.0}{\rmdefault}{\mddefault}{\updefault}\special{ps: gsave 0 0 0 setrgbcolor}$R_2$\special{ps: grestore}}}}
\put(10426,-886){\makebox(0,0)[lb]{\smash{\SetFigFont{5}{6.0}{\rmdefault}{\mddefault}{\updefault}\special{ps: gsave 0 0 0 setrgbcolor}$T_2$\special{ps: grestore}}}}
\put(7801,-886){\makebox(0,0)[lb]{\smash{\SetFigFont{5}{6.0}{\rmdefault}{\mddefault}{\updefault}\special{ps: gsave 0 0 0 setrgbcolor}$R_2$\special{ps: grestore}}}}
\put(6076,-886){\makebox(0,0)[lb]{\smash{\SetFigFont{5}{6.0}{\rmdefault}{\mddefault}{\updefault}\special{ps: gsave 0 0 0 setrgbcolor}$T_2$\special{ps: grestore}}}}
\put(10412,-2104){\makebox(0,0)[lb]{\smash{\SetFigFont{5}{6.0}{\rmdefault}{\mddefault}{\updefault}\special{ps: gsave 0 0 0 setrgbcolor}$W_{11}$\special{ps: grestore}}}}
\put(9385,-164){\makebox(0,0)[lb]{\smash{\SetFigFont{5}{6.0}{\rmdefault}{\mddefault}{\updefault}\special{ps: gsave 0 0 0 setrgbcolor}$W_{11}$\special{ps: grestore}}}}
\put(9385,-1362){\makebox(0,0)[lb]{\smash{\SetFigFont{5}{6.0}{\rmdefault}{\mddefault}{\updefault}\special{ps: gsave 0 0 0 setrgbcolor}$W_{22}$\special{ps: grestore}}}}
\put(12924,-164){\makebox(0,0)[lb]{\smash{\SetFigFont{5}{6.0}{\rmdefault}{\mddefault}{\updefault}\special{ps: gsave 0 0 0 setrgbcolor}$\hat{W}_{11}$\special{ps: grestore}}}}
\put(12924,-1362){\makebox(0,0)[lb]{\smash{\SetFigFont{5}{6.0}{\rmdefault}{\mddefault}{\updefault}\special{ps: gsave 0 0 0 setrgbcolor}$ \hat{W}_{22}$\special{ps: grestore}}}}
\put(1574,-2104){\makebox(0,0)[lb]{\smash{\SetFigFont{5}{6.0}{\rmdefault}{\mddefault}{\updefault}\special{ps: gsave 0 0 0 setrgbcolor}$W_{11}$\special{ps: grestore}}}}
\put(547,-164){\makebox(0,0)[lb]{\smash{\SetFigFont{5}{6.0}{\rmdefault}{\mddefault}{\updefault}\special{ps: gsave 0 0 0 setrgbcolor}$W_{11}$\special{ps: grestore}}}}
\end{picture}

%% file: intcogfull.pstex_t
\begin{picture}(0,0)%
\includegraphics{intcogfull.pstex}%
\end{picture}%
\setlength{\unitlength}{2408sp}%
\begingroup\makeatletter\ifx\SetFigFont\undefined%
\gdef\SetFigFont#1#2#3#4#5{%
  \reset@font\fontsize{#1}{#2pt}%
  \fontfamily{#3}\fontseries{#4}\fontshape{#5}%
  \selectfont}%
\fi\endgroup%
\begin{picture}(13242,3180)(421,-2386)
\put(4086,-164){\makebox(0,0)[lb]{\smash{\SetFigFont{5}{6.0}{\rmdefault}{\mddefault}{\updefault}\special{ps: gsave 0 0 0 setrgbcolor}$\hat{W}_{11}$\special{ps: grestore}}}}
\put(4086,-1362){\makebox(0,0)[lb]{\smash{\SetFigFont{5}{6.0}{\rmdefault}{\mddefault}{\updefault}\special{ps: gsave 0 0 0 setrgbcolor}$ \hat{W}_{22}$\special{ps: grestore}}}}
\put(7774,-2047){\makebox(0,0)[lb]{\smash{\SetFigFont{5}{6.0}{\rmdefault}{\mddefault}{\updefault}\special{ps: gsave 0 0 0 setrgbcolor}$W_{11}$\special{ps: grestore}}}}
\put(5035,-164){\makebox(0,0)[lb]{\smash{\SetFigFont{5}{6.0}{\rmdefault}{\mddefault}{\updefault}\special{ps: gsave 0 0 0 setrgbcolor}$W_{11}$\special{ps: grestore}}}}
\put(5035,-1362){\makebox(0,0)[lb]{\smash{\SetFigFont{5}{6.0}{\rmdefault}{\mddefault}{\updefault}\special{ps: gsave 0 0 0 setrgbcolor}$W_{22}$\special{ps: grestore}}}}
\put(8574,-164){\makebox(0,0)[lb]{\smash{\SetFigFont{5}{6.0}{\rmdefault}{\mddefault}{\updefault}\special{ps: gsave 0 0 0 setrgbcolor}$\hat{W}_{11}$\special{ps: grestore}}}}
\put(8574,-1362){\makebox(0,0)[lb]{\smash{\SetFigFont{5}{6.0}{\rmdefault}{\mddefault}{\updefault}\special{ps: gsave 0 0 0 setrgbcolor}$ \hat{W}_{22}$\special{ps: grestore}}}}
\put(547,-1362){\makebox(0,0)[lb]{\smash{\SetFigFont{5}{6.0}{\rmdefault}{\mddefault}{\updefault}\special{ps: gsave 0 0 0 setrgbcolor}$W_{22}$\special{ps: grestore}}}}
\put(11401,-2311){\makebox(0,0)[lb]{\smash{\SetFigFont{7}{8.4}{\rmdefault}{\mddefault}{\updefault}\special{ps: gsave 0 0 0 setrgbcolor}(c)\special{ps: grestore}}}}
\put(6976,-2311){\makebox(0,0)[lb]{\smash{\SetFigFont{7}{8.4}{\rmdefault}{\mddefault}{\updefault}\special{ps: gsave 0 0 0 setrgbcolor}(b)\special{ps: grestore}}}}
\put(2401,-2386){\makebox(0,0)[lb]{\smash{\SetFigFont{7}{8.4}{\rmdefault}{\mddefault}{\updefault}\special{ps: gsave 0 0 0 setrgbcolor}(a)\special{ps: grestore}}}}
\put(12076,689){\makebox(0,0)[lb]{\smash{\SetFigFont{5}{6.0}{\rmdefault}{\mddefault}{\updefault}\special{ps: gsave 0 0 0 setrgbcolor}$W_{22}$\special{ps: grestore}}}}
\put(7726,689){\makebox(0,0)[lb]{\smash{\SetFigFont{5}{6.0}{\rmdefault}{\mddefault}{\updefault}\special{ps: gsave 0 0 0 setrgbcolor}$W_{22}$\special{ps: grestore}}}}
\put(1576,689){\makebox(0,0)[lb]{\smash{\SetFigFont{5}{6.0}{\rmdefault}{\mddefault}{\updefault}\special{ps: gsave 0 0 0 setrgbcolor}$W_{22}$\special{ps: grestore}}}}
\put(12151,-511){\makebox(0,0)[lb]{\smash{\SetFigFont{5}{6.0}{\rmdefault}{\mddefault}{\updefault}\special{ps: gsave 0 0 0 setrgbcolor}$R_1$\special{ps: grestore}}}}
\put(10426,-511){\makebox(0,0)[lb]{\smash{\SetFigFont{5}{6.0}{\rmdefault}{\mddefault}{\updefault}\special{ps: gsave 0 0 0 setrgbcolor}$T_1$\special{ps: grestore}}}}
\put(7876,-511){\makebox(0,0)[lb]{\smash{\SetFigFont{5}{6.0}{\rmdefault}{\mddefault}{\updefault}\special{ps: gsave 0 0 0 setrgbcolor}$R_1$\special{ps: grestore}}}}
\put(6076,-511){\makebox(0,0)[lb]{\smash{\SetFigFont{5}{6.0}{\rmdefault}{\mddefault}{\updefault}\special{ps: gsave 0 0 0 setrgbcolor}$T_1$\special{ps: grestore}}}}
\put(1576,-511){\makebox(0,0)[lb]{\smash{\SetFigFont{5}{6.0}{\rmdefault}{\mddefault}{\updefault}\special{ps: gsave 0 0 0 setrgbcolor}$T_1$\special{ps: grestore}}}}
\put(3376,-511){\makebox(0,0)[lb]{\smash{\SetFigFont{5}{6.0}{\rmdefault}{\mddefault}{\updefault}\special{ps: gsave 0 0 0 setrgbcolor}$R_1$\special{ps: grestore}}}}
\put(3376,-886){\makebox(0,0)[lb]{\smash{\SetFigFont{5}{6.0}{\rmdefault}{\mddefault}{\updefault}\special{ps: gsave 0 0 0 setrgbcolor}$R_2$\special{ps: grestore}}}}
\put(1576,-886){\makebox(0,0)[lb]{\smash{\SetFigFont{5}{6.0}{\rmdefault}{\mddefault}{\updefault}\special{ps: gsave 0 0 0 setrgbcolor}$T_2$\special{ps: grestore}}}}
\put(12151,-886){\makebox(0,0)[lb]{\smash{\SetFigFont{5}{6.0}{\rmdefault}{\mddefault}{\updefault}\special{ps: gsave 0 0 0 setrgbcolor}$R_2$\special{ps: grestore}}}}
\put(10426,-886){\makebox(0,0)[lb]{\smash{\SetFigFont{5}{6.0}{\rmdefault}{\mddefault}{\updefault}\special{ps: gsave 0 0 0 setrgbcolor}$T_2$\special{ps: grestore}}}}
\put(7801,-886){\makebox(0,0)[lb]{\smash{\SetFigFont{5}{6.0}{\rmdefault}{\mddefault}{\updefault}\special{ps: gsave 0 0 0 setrgbcolor}$R_2$\special{ps: grestore}}}}
\put(6076,-886){\makebox(0,0)[lb]{\smash{\SetFigFont{5}{6.0}{\rmdefault}{\mddefault}{\updefault}\special{ps: gsave 0 0 0 setrgbcolor}$T_2$\special{ps: grestore}}}}
\put(10412,-2104){\makebox(0,0)[lb]{\smash{\SetFigFont{5}{6.0}{\rmdefault}{\mddefault}{\updefault}\special{ps: gsave 0 0 0 setrgbcolor}$W_{11}$\special{ps: grestore}}}}
\put(9385,-164){\makebox(0,0)[lb]{\smash{\SetFigFont{5}{6.0}{\rmdefault}{\mddefault}{\updefault}\special{ps: gsave 0 0 0 setrgbcolor}$W_{11}$\special{ps: grestore}}}}
\put(9385,-1362){\makebox(0,0)[lb]{\smash{\SetFigFont{5}{6.0}{\rmdefault}{\mddefault}{\updefault}\special{ps: gsave 0 0 0 setrgbcolor}$W_{22}$\special{ps: grestore}}}}
\put(12924,-164){\makebox(0,0)[lb]{\smash{\SetFigFont{5}{6.0}{\rmdefault}{\mddefault}{\updefault}\special{ps: gsave 0 0 0 setrgbcolor}$\hat{W}_{11}$\special{ps: grestore}}}}
\put(12924,-1362){\makebox(0,0)[lb]{\smash{\SetFigFont{5}{6.0}{\rmdefault}{\mddefault}{\updefault}\special{ps: gsave 0 0 0 setrgbcolor}$ \hat{W}_{22}$\special{ps: grestore}}}}
\put(1574,-2104){\makebox(0,0)[lb]{\smash{\SetFigFont{5}{6.0}{\rmdefault}{\mddefault}{\updefault}\special{ps: gsave 0 0 0 setrgbcolor}$W_{11}$\special{ps: grestore}}}}
\put(547,-164){\makebox(0,0)[lb]{\smash{\SetFigFont{5}{6.0}{\rmdefault}{\mddefault}{\updefault}\special{ps: gsave 0 0 0 setrgbcolor}$W_{11}$\special{ps: grestore}}}}
\end{picture}

%% file: cogxfull.pstex_t
\begin{picture}(0,0)%
\includegraphics{cogxfull.pstex}%
\end{picture}%
\setlength{\unitlength}{2408sp}%
\begingroup\makeatletter\ifx\SetFigFont\undefined%
\gdef\SetFigFont#1#2#3#4#5{%
  \reset@font\fontsize{#1}{#2pt}%
  \fontfamily{#3}\fontseries{#4}\fontshape{#5}%
  \selectfont}%
\fi\endgroup%
\begin{picture}(8784,2712)(421,-2311)
\put(4051,-1486){\makebox(0,0)[lb]{\smash{\SetFigFont{5}{6.0}{\rmdefault}{\mddefault}{\updefault}\special{ps: gsave 0 0 0 setrgbcolor}$ \hat{W}_{22}$\special{ps: grestore}}}}
\put(526,-286){\makebox(0,0)[lb]{\smash{\SetFigFont{5}{6.0}{\rmdefault}{\mddefault}{\updefault}\special{ps: gsave 0 0 0 setrgbcolor}$W_{21}$\special{ps: grestore}}}}
\put(526,-1186){\makebox(0,0)[lb]{\smash{\SetFigFont{5}{6.0}{\rmdefault}{\mddefault}{\updefault}\special{ps: gsave 0 0 0 setrgbcolor}$W_{12}$\special{ps: grestore}}}}
\put(526,-1486){\makebox(0,0)[lb]{\smash{\SetFigFont{5}{6.0}{\rmdefault}{\mddefault}{\updefault}\special{ps: gsave 0 0 0 setrgbcolor}$W_{22}$\special{ps: grestore}}}}
\put(1576,-511){\makebox(0,0)[lb]{\smash{\SetFigFont{5}{6.0}{\rmdefault}{\mddefault}{\updefault}\special{ps: gsave 0 0 0 setrgbcolor}$T_1$\special{ps: grestore}}}}
\put(3376,-511){\makebox(0,0)[lb]{\smash{\SetFigFont{5}{6.0}{\rmdefault}{\mddefault}{\updefault}\special{ps: gsave 0 0 0 setrgbcolor}$R_1$\special{ps: grestore}}}}
\put(3376,-886){\makebox(0,0)[lb]{\smash{\SetFigFont{5}{6.0}{\rmdefault}{\mddefault}{\updefault}\special{ps: gsave 0 0 0 setrgbcolor}$R_2$\special{ps: grestore}}}}
\put(1576,-886){\makebox(0,0)[lb]{\smash{\SetFigFont{5}{6.0}{\rmdefault}{\mddefault}{\updefault}\special{ps: gsave 0 0 0 setrgbcolor}$T_2$\special{ps: grestore}}}}
\put(6226,-886){\makebox(0,0)[lb]{\smash{\SetFigFont{5}{6.0}{\rmdefault}{\mddefault}{\updefault}\special{ps: gsave 0 0 0 setrgbcolor}$T_2$\special{ps: grestore}}}}
\put(7201,-2311){\makebox(0,0)[lb]{\smash{\SetFigFont{7}{8.4}{\rmdefault}{\mddefault}{\updefault}\special{ps: gsave 0 0 0 setrgbcolor}(b)\special{ps: grestore}}}}
\put(2401,-2311){\makebox(0,0)[lb]{\smash{\SetFigFont{7}{8.4}{\rmdefault}{\mddefault}{\updefault}\special{ps: gsave 0 0 0 setrgbcolor}(a)\special{ps: grestore}}}}
\put(8701,-1186){\makebox(0,0)[lb]{\smash{\SetFigFont{5}{6.0}{\rmdefault}{\mddefault}{\updefault}\special{ps: gsave 0 0 0 setrgbcolor}$ \hat{W}_{21}$\special{ps: grestore}}}}
\put(4051,-61){\makebox(0,0)[lb]{\smash{\SetFigFont{5}{6.0}{\rmdefault}{\mddefault}{\updefault}\special{ps: gsave 0 0 0 setrgbcolor}$\hat{W}_{11}$\special{ps: grestore}}}}
\put(5176,-286){\makebox(0,0)[lb]{\smash{\SetFigFont{5}{6.0}{\rmdefault}{\mddefault}{\updefault}\special{ps: gsave 0 0 0 setrgbcolor}$W_{21}$\special{ps: grestore}}}}
\put(526, 14){\makebox(0,0)[lb]{\smash{\SetFigFont{5}{6.0}{\rmdefault}{\mddefault}{\updefault}\special{ps: gsave 0 0 0 setrgbcolor}$W_{11}$\special{ps: grestore}}}}
\put(1576,-2086){\makebox(0,0)[lb]{\smash{\SetFigFont{5}{6.0}{\rmdefault}{\mddefault}{\updefault}\special{ps: gsave 0 0 0 setrgbcolor}$W_{11}$\special{ps: grestore}}}}
\put(8701,-286){\makebox(0,0)[lb]{\smash{\SetFigFont{5}{6.0}{\rmdefault}{\mddefault}{\updefault}\special{ps: gsave 0 0 0 setrgbcolor}$\hat{W}_{12}$\special{ps: grestore}}}}
\put(8701, 14){\makebox(0,0)[lb]{\smash{\SetFigFont{5}{6.0}{\rmdefault}{\mddefault}{\updefault}\special{ps: gsave 0 0 0 setrgbcolor}$\hat{W}_{11}$\special{ps: grestore}}}}
\put(4051,-286){\makebox(0,0)[lb]{\smash{\SetFigFont{5}{6.0}{\rmdefault}{\mddefault}{\updefault}\special{ps: gsave 0 0 0 setrgbcolor}$\hat{W}_{12}$\special{ps: grestore}}}}
\put(4051,-1186){\makebox(0,0)[lb]{\smash{\SetFigFont{5}{6.0}{\rmdefault}{\mddefault}{\updefault}\special{ps: gsave 0 0 0 setrgbcolor}$ \hat{W}_{21}$\special{ps: grestore}}}}
\put(8701,-1486){\makebox(0,0)[lb]{\smash{\SetFigFont{5}{6.0}{\rmdefault}{\mddefault}{\updefault}\special{ps: gsave 0 0 0 setrgbcolor}$ \hat{W}_{22}$\special{ps: grestore}}}}
\put(5176,-1186){\makebox(0,0)[lb]{\smash{\SetFigFont{5}{6.0}{\rmdefault}{\mddefault}{\updefault}\special{ps: gsave 0 0 0 setrgbcolor}$W_{12}$\special{ps: grestore}}}}
\put(7951,-2161){\makebox(0,0)[lb]{\smash{\SetFigFont{5}{6.0}{\rmdefault}{\mddefault}{\updefault}\special{ps: gsave 0 0 0 setrgbcolor}$W_{11}$\special{ps: grestore}}}}
\put(5176,-1486){\makebox(0,0)[lb]{\smash{\SetFigFont{5}{6.0}{\rmdefault}{\mddefault}{\updefault}\special{ps: gsave 0 0 0 setrgbcolor}$W_{22}$\special{ps: grestore}}}}
\put(5176,-61){\makebox(0,0)[lb]{\smash{\SetFigFont{5}{6.0}{\rmdefault}{\mddefault}{\updefault}\special{ps: gsave 0 0 0 setrgbcolor}$W_{11}$\special{ps: grestore}}}}
\put(6226,-511){\makebox(0,0)[lb]{\smash{\SetFigFont{5}{6.0}{\rmdefault}{\mddefault}{\updefault}\special{ps: gsave 0 0 0 setrgbcolor}$T_1$\special{ps: grestore}}}}
\put(8026,-511){\makebox(0,0)[lb]{\smash{\SetFigFont{5}{6.0}{\rmdefault}{\mddefault}{\updefault}\special{ps: gsave 0 0 0 setrgbcolor}$R_1$\special{ps: grestore}}}}
\put(8026,-886){\makebox(0,0)[lb]{\smash{\SetFigFont{5}{6.0}{\rmdefault}{\mddefault}{\updefault}\special{ps: gsave 0 0 0 setrgbcolor}$R_2$\special{ps: grestore}}}}
\end{picture}

%% file: zint.pstex_t
\begin{picture}(0,0)%
\includegraphics{zint.pstex}%
\end{picture}%
\setlength{\unitlength}{2408sp}%
\begingroup\makeatletter\ifx\SetFigFont\undefined%
\gdef\SetFigFont#1#2#3#4#5{%
  \reset@font\fontsize{#1}{#2pt}%
  \fontfamily{#3}\fontseries{#4}\fontshape{#5}%
  \selectfont}%
\fi\endgroup%
\begin{picture}(10983,3108)(118,-2236)
\put(4951,239){\makebox(0,0)[lb]{\smash{\SetFigFont{7}{8.4}{\rmdefault}{\mddefault}{\updefault}\special{ps: gsave 0 0 0 setrgbcolor}$\hat{W}_{11}$\special{ps: grestore}}}}
\put(4951,-61){\makebox(0,0)[lb]{\smash{\SetFigFont{7}{8.4}{\rmdefault}{\mddefault}{\updefault}\special{ps: gsave 0 0 0 setrgbcolor}$\hat{W}_{12}$\special{ps: grestore}}}}
\put(4951,-1636){\makebox(0,0)[lb]{\smash{\SetFigFont{7}{8.4}{\rmdefault}{\mddefault}{\updefault}\special{ps: gsave 0 0 0 setrgbcolor}$\hat{W}_{22}$\special{ps: grestore}}}}
\put(7726,-2236){\makebox(0,0)[lb]{\smash{\SetFigFont{7}{8.4}{\rmdefault}{\mddefault}{\updefault}\special{ps: gsave 0 0 0 setrgbcolor}(b) The MIMO Interference Channel\special{ps: grestore}}}}
\put(1576,-2236){\makebox(0,0)[lb]{\smash{\SetFigFont{7}{8.4}{\rmdefault}{\mddefault}{\updefault}\special{ps: gsave 0 0 0 setrgbcolor}(a) The MIMO $Z$ Channel\special{ps: grestore}}}}
\put(301,-1561){\makebox(0,0)[lb]{\smash{\SetFigFont{7}{8.4}{\rmdefault}{\mddefault}{\updefault}\special{ps: gsave 0 0 0 setrgbcolor}$W_{22}$\special{ps: grestore}}}}
\put(301,-1261){\makebox(0,0)[lb]{\smash{\SetFigFont{7}{8.4}{\rmdefault}{\mddefault}{\updefault}\special{ps: gsave 0 0 0 setrgbcolor}$W_{12}$\special{ps: grestore}}}}
\put(301,314){\makebox(0,0)[lb]{\smash{\SetFigFont{7}{8.4}{\rmdefault}{\mddefault}{\updefault}\special{ps: gsave 0 0 0 setrgbcolor}$W_{11}$\special{ps: grestore}}}}
\put(3976,-961){\makebox(0,0)[lb]{\smash{\SetFigFont{7}{8.4}{\rmdefault}{\mddefault}{\updefault}\special{ps: gsave 0 0 0 setrgbcolor}$R_2$\special{ps: grestore}}}}
\put(1651,-961){\makebox(0,0)[lb]{\smash{\SetFigFont{7}{8.4}{\rmdefault}{\mddefault}{\updefault}\special{ps: gsave 0 0 0 setrgbcolor}$T_2$\special{ps: grestore}}}}
\put(1651,614){\makebox(0,0)[lb]{\smash{\SetFigFont{7}{8.4}{\rmdefault}{\mddefault}{\updefault}\special{ps: gsave 0 0 0 setrgbcolor}$T_1$\special{ps: grestore}}}}
\put(3976,614){\makebox(0,0)[lb]{\smash{\SetFigFont{7}{8.4}{\rmdefault}{\mddefault}{\updefault}\special{ps: gsave 0 0 0 setrgbcolor}$R_1$\special{ps: grestore}}}}
\put(11101,-1636){\makebox(0,0)[lb]{\smash{\SetFigFont{7}{8.4}{\rmdefault}{\mddefault}{\updefault}\special{ps: gsave 0 0 0 setrgbcolor}$\hat{W}_{22}$\special{ps: grestore}}}}
\put(11101,239){\makebox(0,0)[lb]{\smash{\SetFigFont{7}{8.4}{\rmdefault}{\mddefault}{\updefault}\special{ps: gsave 0 0 0 setrgbcolor}$\hat{W}_{11}$\special{ps: grestore}}}}
\put(6451,-1561){\makebox(0,0)[lb]{\smash{\SetFigFont{7}{8.4}{\rmdefault}{\mddefault}{\updefault}\special{ps: gsave 0 0 0 setrgbcolor}$W_{22}$\special{ps: grestore}}}}
\put(6451,314){\makebox(0,0)[lb]{\smash{\SetFigFont{7}{8.4}{\rmdefault}{\mddefault}{\updefault}\special{ps: gsave 0 0 0 setrgbcolor}$W_{11}$\special{ps: grestore}}}}
\put(10126,-961){\makebox(0,0)[lb]{\smash{\SetFigFont{7}{8.4}{\rmdefault}{\mddefault}{\updefault}\special{ps: gsave 0 0 0 setrgbcolor}$R_2$\special{ps: grestore}}}}
\put(7801,-961){\makebox(0,0)[lb]{\smash{\SetFigFont{7}{8.4}{\rmdefault}{\mddefault}{\updefault}\special{ps: gsave 0 0 0 setrgbcolor}$T_2$\special{ps: grestore}}}}
\put(7801,614){\makebox(0,0)[lb]{\smash{\SetFigFont{7}{8.4}{\rmdefault}{\mddefault}{\updefault}\special{ps: gsave 0 0 0 setrgbcolor}$T_1$\special{ps: grestore}}}}
\put(10126,614){\makebox(0,0)[lb]{\smash{\SetFigFont{7}{8.4}{\rmdefault}{\mddefault}{\updefault}\special{ps: gsave 0 0 0 setrgbcolor}$R_1$\special{ps: grestore}}}}
\end{picture}

%% file: xscene5.pstex_t
\begin{picture}(0,0)%
\includegraphics{xscene5.pstex}%
\end{picture}%
\setlength{\unitlength}{2408sp}%
\begingroup\makeatletter\ifx\SetFigFont\undefined%
\gdef\SetFigFont#1#2#3#4#5{%
  \reset@font\fontsize{#1}{#2pt}%
  \fontfamily{#3}\fontseries{#4}\fontshape{#5}%
  \selectfont}%
\fi\endgroup%
\begin{picture}(11691,3166)(418,-2744)
\put(6826,-136){\makebox(0,0)[lb]{\smash{\SetFigFont{7}{8.4}{\rmdefault}{\mddefault}{\updefault}\special{ps: gsave 0 0 0 setrgbcolor}$W_1$\special{ps: grestore}}}}
\put(6826,-1186){\makebox(0,0)[lb]{\smash{\SetFigFont{7}{8.4}{\rmdefault}{\mddefault}{\updefault}\special{ps: gsave 0 0 0 setrgbcolor}$W_0$\special{ps: grestore}}}}
\put(8176,164){\makebox(0,0)[lb]{\smash{\SetFigFont{7}{8.4}{\rmdefault}{\mddefault}{\updefault}\special{ps: gsave 0 0 0 setrgbcolor}$T_1$\special{ps: grestore}}}}
\put(10501,164){\makebox(0,0)[lb]{\smash{\SetFigFont{7}{8.4}{\rmdefault}{\mddefault}{\updefault}\special{ps: gsave 0 0 0 setrgbcolor}$R_1$\special{ps: grestore}}}}
\put(6826,-2011){\makebox(0,0)[lb]{\smash{\SetFigFont{7}{8.4}{\rmdefault}{\mddefault}{\updefault}\special{ps: gsave 0 0 0 setrgbcolor}$W_2$\special{ps: grestore}}}}
\put(7726,-2686){\makebox(0,0)[lb]{\smash{\SetFigFont{7}{8.4}{\rmdefault}{\mddefault}{\updefault}\special{ps: gsave 0 0 0 setrgbcolor}(b) Equivalent channel model with $3$ messages.\special{ps: grestore}}}}
\put(1351,-2686){\makebox(0,0)[lb]{\smash{\SetFigFont{7}{8.4}{\rmdefault}{\mddefault}{\updefault}\special{ps: gsave 0 0 0 setrgbcolor}(a) MIMO X channel with cognitive transmiter\special{ps: grestore}}}}
\put(601,-136){\makebox(0,0)[lb]{\smash{\SetFigFont{7}{8.4}{\rmdefault}{\mddefault}{\updefault}\special{ps: gsave 0 0 0 setrgbcolor}$W_{21}$\special{ps: grestore}}}}
\put(601,-511){\makebox(0,0)[lb]{\smash{\SetFigFont{7}{8.4}{\rmdefault}{\mddefault}{\updefault}\special{ps: gsave 0 0 0 setrgbcolor}$W_{11}$\special{ps: grestore}}}}
\put(5251,-1936){\makebox(0,0)[lb]{\smash{\SetFigFont{7}{8.4}{\rmdefault}{\mddefault}{\updefault}\special{ps: gsave 0 0 0 setrgbcolor}$\hat{W}_{21}, \hat{W}_{22}$\special{ps: grestore}}}}
\put(11476,-1936){\makebox(0,0)[lb]{\smash{\SetFigFont{7}{8.4}{\rmdefault}{\mddefault}{\updefault}\special{ps: gsave 0 0 0 setrgbcolor}$\hat{W}_{1}, \hat{W}_{2}$\special{ps: grestore}}}}
\put(6076,-1111){\makebox(0,0)[lb]{\smash{\SetFigFont{7}{8.4}{\rmdefault}{\mddefault}{\updefault}\special{ps: gsave 0 0 0 setrgbcolor}$\equiv$\special{ps: grestore}}}}
\put(601,-2236){\makebox(0,0)[lb]{\smash{\SetFigFont{7}{8.4}{\rmdefault}{\mddefault}{\updefault}\special{ps: gsave 0 0 0 setrgbcolor}$W_{22}$\special{ps: grestore}}}}
\put(601,-2011){\makebox(0,0)[lb]{\smash{\SetFigFont{7}{8.4}{\rmdefault}{\mddefault}{\updefault}\special{ps: gsave 0 0 0 setrgbcolor}$W_{12}$\special{ps: grestore}}}}
\put(5251,-361){\makebox(0,0)[lb]{\smash{\SetFigFont{7}{8.4}{\rmdefault}{\mddefault}{\updefault}\special{ps: gsave 0 0 0 setrgbcolor}$\hat{W}_{11},\hat{W}_{12}$\special{ps: grestore}}}}
\put(4276,-1411){\makebox(0,0)[lb]{\smash{\SetFigFont{7}{8.4}{\rmdefault}{\mddefault}{\updefault}\special{ps: gsave 0 0 0 setrgbcolor}$R_2$\special{ps: grestore}}}}
\put(1951,-1411){\makebox(0,0)[lb]{\smash{\SetFigFont{7}{8.4}{\rmdefault}{\mddefault}{\updefault}\special{ps: gsave 0 0 0 setrgbcolor}$T_2$\special{ps: grestore}}}}
\put(1951,164){\makebox(0,0)[lb]{\smash{\SetFigFont{7}{8.4}{\rmdefault}{\mddefault}{\updefault}\special{ps: gsave 0 0 0 setrgbcolor}$T_1$\special{ps: grestore}}}}
\put(4276,164){\makebox(0,0)[lb]{\smash{\SetFigFont{7}{8.4}{\rmdefault}{\mddefault}{\updefault}\special{ps: gsave 0 0 0 setrgbcolor}$R_1$\special{ps: grestore}}}}
\put(10501,-1411){\makebox(0,0)[lb]{\smash{\SetFigFont{7}{8.4}{\rmdefault}{\mddefault}{\updefault}\special{ps: gsave 0 0 0 setrgbcolor}$R_2$\special{ps: grestore}}}}
\put(8176,-1411){\makebox(0,0)[lb]{\smash{\SetFigFont{7}{8.4}{\rmdefault}{\mddefault}{\updefault}\special{ps: gsave 0 0 0 setrgbcolor}$T_2$\special{ps: grestore}}}}
\put(11476,-361){\makebox(0,0)[lb]{\smash{\SetFigFont{7}{8.4}{\rmdefault}{\mddefault}{\updefault}\special{ps: gsave 0 0 0 setrgbcolor}$\hat{W}_0$\special{ps: grestore}}}}
\end{picture}

%% file: cogxreceiver.pstex_t
\begin{picture}(0,0)%
\includegraphics{cogxreceiver.pstex}%
\end{picture}%
\setlength{\unitlength}{2408sp}%
\begingroup\makeatletter\ifx\SetFigFont\undefined%
\gdef\SetFigFont#1#2#3#4#5{%
  \reset@font\fontsize{#1}{#2pt}%
  \fontfamily{#3}\fontseries{#4}\fontshape{#5}%
  \selectfont}%
\fi\endgroup%
\begin{picture}(13434,2845)(421,-2444)
\put(6226,-886){\makebox(0,0)[lb]{\smash{\SetFigFont{5}{6.0}{\rmdefault}{\mddefault}{\updefault}\special{ps: gsave 0 0 0 setrgbcolor}$T_2$\special{ps: grestore}}}}
\put(4051,-1486){\makebox(0,0)[lb]{\smash{\SetFigFont{5}{6.0}{\rmdefault}{\mddefault}{\updefault}\special{ps: gsave 0 0 0 setrgbcolor}$ \hat{W}_{22}$\special{ps: grestore}}}}
\put(526,-286){\makebox(0,0)[lb]{\smash{\SetFigFont{5}{6.0}{\rmdefault}{\mddefault}{\updefault}\special{ps: gsave 0 0 0 setrgbcolor}$W_{21}$\special{ps: grestore}}}}
\put(526,-1186){\makebox(0,0)[lb]{\smash{\SetFigFont{5}{6.0}{\rmdefault}{\mddefault}{\updefault}\special{ps: gsave 0 0 0 setrgbcolor}$W_{12}$\special{ps: grestore}}}}
\put(526,-1486){\makebox(0,0)[lb]{\smash{\SetFigFont{5}{6.0}{\rmdefault}{\mddefault}{\updefault}\special{ps: gsave 0 0 0 setrgbcolor}$W_{22}$\special{ps: grestore}}}}
\put(1576,-511){\makebox(0,0)[lb]{\smash{\SetFigFont{5}{6.0}{\rmdefault}{\mddefault}{\updefault}\special{ps: gsave 0 0 0 setrgbcolor}$T_1$\special{ps: grestore}}}}
\put(3376,-511){\makebox(0,0)[lb]{\smash{\SetFigFont{5}{6.0}{\rmdefault}{\mddefault}{\updefault}\special{ps: gsave 0 0 0 setrgbcolor}$R_1$\special{ps: grestore}}}}
\put(3376,-886){\makebox(0,0)[lb]{\smash{\SetFigFont{5}{6.0}{\rmdefault}{\mddefault}{\updefault}\special{ps: gsave 0 0 0 setrgbcolor}$R_2$\special{ps: grestore}}}}
\put(1576,-886){\makebox(0,0)[lb]{\smash{\SetFigFont{5}{6.0}{\rmdefault}{\mddefault}{\updefault}\special{ps: gsave 0 0 0 setrgbcolor}$T_2$\special{ps: grestore}}}}
\put(8026,-886){\makebox(0,0)[lb]{\smash{\SetFigFont{5}{6.0}{\rmdefault}{\mddefault}{\updefault}\special{ps: gsave 0 0 0 setrgbcolor}$R_2$\special{ps: grestore}}}}
\put(12976,-2161){\makebox(0,0)[lb]{\smash{\SetFigFont{5}{6.0}{\rmdefault}{\mddefault}{\updefault}\special{ps: gsave 0 0 0 setrgbcolor}$W_{12}$\special{ps: grestore}}}}
\put(12451,-2161){\makebox(0,0)[lb]{\smash{\SetFigFont{5}{6.0}{\rmdefault}{\mddefault}{\updefault}\special{ps: gsave 0 0 0 setrgbcolor}$W_{11}$\special{ps: grestore}}}}
\put(11101,-2386){\makebox(0,0)[lb]{\smash{\SetFigFont{7}{8.4}{\rmdefault}{\mddefault}{\updefault}\special{ps: gsave 0 0 0 setrgbcolor}(c) $W_{22}=\phi$\special{ps: grestore}}}}
\put(6376,-2386){\makebox(0,0)[lb]{\smash{\SetFigFont{7}{8.4}{\rmdefault}{\mddefault}{\updefault}\special{ps: gsave 0 0 0 setrgbcolor}(b) $W_{21}=\phi$\special{ps: grestore}}}}
\put(1726,-2386){\makebox(0,0)[lb]{\smash{\SetFigFont{7}{8.4}{\rmdefault}{\mddefault}{\updefault}\special{ps: gsave 0 0 0 setrgbcolor}(a) $W_{11}=\phi$\special{ps: grestore}}}}
\put(13351,-286){\makebox(0,0)[lb]{\smash{\SetFigFont{5}{6.0}{\rmdefault}{\mddefault}{\updefault}\special{ps: gsave 0 0 0 setrgbcolor}$\hat{W}_{12}$\special{ps: grestore}}}}
\put(13351,-1186){\makebox(0,0)[lb]{\smash{\SetFigFont{5}{6.0}{\rmdefault}{\mddefault}{\updefault}\special{ps: gsave 0 0 0 setrgbcolor}$ \hat{W}_{21}$\special{ps: grestore}}}}
\put(8701,-286){\makebox(0,0)[lb]{\smash{\SetFigFont{5}{6.0}{\rmdefault}{\mddefault}{\updefault}\special{ps: gsave 0 0 0 setrgbcolor}$\hat{W}_{12}$\special{ps: grestore}}}}
\put(8701, 14){\makebox(0,0)[lb]{\smash{\SetFigFont{5}{6.0}{\rmdefault}{\mddefault}{\updefault}\special{ps: gsave 0 0 0 setrgbcolor}$\hat{W}_{11}$\special{ps: grestore}}}}
\put(4051,-286){\makebox(0,0)[lb]{\smash{\SetFigFont{5}{6.0}{\rmdefault}{\mddefault}{\updefault}\special{ps: gsave 0 0 0 setrgbcolor}$\hat{W}_{12}$\special{ps: grestore}}}}
\put(4051,-1186){\makebox(0,0)[lb]{\smash{\SetFigFont{5}{6.0}{\rmdefault}{\mddefault}{\updefault}\special{ps: gsave 0 0 0 setrgbcolor}$ \hat{W}_{21}$\special{ps: grestore}}}}
\put(13351, 14){\makebox(0,0)[lb]{\smash{\SetFigFont{5}{6.0}{\rmdefault}{\mddefault}{\updefault}\special{ps: gsave 0 0 0 setrgbcolor}$\hat{W}_{11}$\special{ps: grestore}}}}
\put(9826,-286){\makebox(0,0)[lb]{\smash{\SetFigFont{5}{6.0}{\rmdefault}{\mddefault}{\updefault}\special{ps: gsave 0 0 0 setrgbcolor}$W_{21}$\special{ps: grestore}}}}
\put(9826,-1186){\makebox(0,0)[lb]{\smash{\SetFigFont{5}{6.0}{\rmdefault}{\mddefault}{\updefault}\special{ps: gsave 0 0 0 setrgbcolor}$W_{12}$\special{ps: grestore}}}}
\put(9826,-61){\makebox(0,0)[lb]{\smash{\SetFigFont{5}{6.0}{\rmdefault}{\mddefault}{\updefault}\special{ps: gsave 0 0 0 setrgbcolor}$W_{11}$\special{ps: grestore}}}}
\put(10876,-511){\makebox(0,0)[lb]{\smash{\SetFigFont{5}{6.0}{\rmdefault}{\mddefault}{\updefault}\special{ps: gsave 0 0 0 setrgbcolor}$T_1$\special{ps: grestore}}}}
\put(12676,-511){\makebox(0,0)[lb]{\smash{\SetFigFont{5}{6.0}{\rmdefault}{\mddefault}{\updefault}\special{ps: gsave 0 0 0 setrgbcolor}$R_1$\special{ps: grestore}}}}
\put(12676,-886){\makebox(0,0)[lb]{\smash{\SetFigFont{5}{6.0}{\rmdefault}{\mddefault}{\updefault}\special{ps: gsave 0 0 0 setrgbcolor}$R_2$\special{ps: grestore}}}}
\put(10876,-886){\makebox(0,0)[lb]{\smash{\SetFigFont{5}{6.0}{\rmdefault}{\mddefault}{\updefault}\special{ps: gsave 0 0 0 setrgbcolor}$T_2$\special{ps: grestore}}}}
\put(8701,-1486){\makebox(0,0)[lb]{\smash{\SetFigFont{5}{6.0}{\rmdefault}{\mddefault}{\updefault}\special{ps: gsave 0 0 0 setrgbcolor}$ \hat{W}_{22}$\special{ps: grestore}}}}
\put(5176,-1186){\makebox(0,0)[lb]{\smash{\SetFigFont{5}{6.0}{\rmdefault}{\mddefault}{\updefault}\special{ps: gsave 0 0 0 setrgbcolor}$W_{12}$\special{ps: grestore}}}}
\put(7951,-2161){\makebox(0,0)[lb]{\smash{\SetFigFont{5}{6.0}{\rmdefault}{\mddefault}{\updefault}\special{ps: gsave 0 0 0 setrgbcolor}$W_{11}$\special{ps: grestore}}}}
\put(5176,-1486){\makebox(0,0)[lb]{\smash{\SetFigFont{5}{6.0}{\rmdefault}{\mddefault}{\updefault}\special{ps: gsave 0 0 0 setrgbcolor}$W_{22}$\special{ps: grestore}}}}
\put(5176,-61){\makebox(0,0)[lb]{\smash{\SetFigFont{5}{6.0}{\rmdefault}{\mddefault}{\updefault}\special{ps: gsave 0 0 0 setrgbcolor}$W_{11}$\special{ps: grestore}}}}
\put(6226,-511){\makebox(0,0)[lb]{\smash{\SetFigFont{5}{6.0}{\rmdefault}{\mddefault}{\updefault}\special{ps: gsave 0 0 0 setrgbcolor}$T_1$\special{ps: grestore}}}}
\put(8026,-511){\makebox(0,0)[lb]{\smash{\SetFigFont{5}{6.0}{\rmdefault}{\mddefault}{\updefault}\special{ps: gsave 0 0 0 setrgbcolor}$R_1$\special{ps: grestore}}}}
\end{picture}